\documentclass[aps,pra,superscriptaddress,twocolumn]{revtex4-1}
\usepackage[utf8]{inputenc}
\usepackage{color}
\usepackage{verbatim}
\usepackage{amsmath}
\usepackage{amssymb}
\usepackage{graphicx}
\usepackage{microtype}
\usepackage[unicode=true,
 bookmarks=false,
 breaklinks=false,pdfborder={0 0 1},backref=false,colorlinks=true]
 {hyperref}
\hypersetup{
 linkcolor=blue,citecolor=blue,urlcolor=blue,linkcolor=blue,citecolor=blue,urlcolor=blue}

\makeatletter
\usepackage{adjustbox}
\definecolor{mygreen}{rgb}{0,0.5,0}
\definecolor{myblue}{rgb}{0,0,0.75}
\definecolor{mymagenta}{cmyk}{0,1,0,0.12}

\newcommand\tab[1][1cm]{\hspace*{#1}}

\newcommand{\diagram}[1]{\adjustimage{raise=-.5\height}{diagrams/#1.pdf}}

\newcommand{\mylabel}[2][0]{%
    \refstepcounter{equation}\label{#2}%
    \marginnote{%
        \ifnum#1>0%
            \addtocounter{equation}{-#1}%
            \newcount\inlinecount%
            \inlinecount=0%
            \loop\ifnum\inlinecount<#1%
            \setbox0\hbox{\theequation}%
            \hskip\wd0,% This % was missing resulting in more space than 1ex
            \hskip0ex%
            \stepcounter{equation}%
            \advance\inlinecount by 1%
            \repeat%
        \fi%
        \hskip-10ex%
        \hspace*{\fill}(\theequation)\tab[8.36ex]%
    }%
}

\renewcommand{\citet}{\cite}

\makeatother

\begin{document}
\title{Diagrammatic technique for simulation of large-scale quantum repeater
networks with dissipating quantum memories}
\author{Viacheslav V. Kuzmin}
\author{Denis V. Vasilyev}
\affiliation{Center for Quantum Physics, University of Innsbruck, 6020 Innsbruck,
Austria}
\affiliation{Institute for Quantum Optics and Quantum Information of the Austrian
Academy of Sciences, 6020 Innsbruck, Austria}
\date{\today}
\begin{abstract}
We present a detailed description of the diagrammatic technique, recently
devised in {[}V.~V.~Kuzmin\emph{~et.~al.}, npj Quantum Information
5, 115 (2019){]}, for semi-analytical description of large-scale quantum-repeater
networks. The technique takes into account all essential experimental
imperfections, including dissipative Liouville dynamics of the network
quantum memories and the classical communication delays. The results
obtained with the semi-analytic method match the exact Monte Carlo
simulations while the required computational resources scale only
linearly with the network size. The presented approach opens new possibilities
for the development and efficient optimization of future quantum networks.
\end{abstract}
\maketitle

\section{Introduction}

\noindent 

Quantum internet~\citep{Wehner2018} aims at generating entangled
states distributed over long distances, enabling such quantum technologies
as secure communication~\citet{Bruss2000}, distributed quantum computing~\citet{Beals2013},
and distributed metrology~\citet{Komar14,Eldredge2016,Proctor2017,Ge2017}.
The major problem for implementation of a quantum internet is the
exponential loss of photons — carriers of the entanglement — with
the increasing length of the lossy channel. Quantum repeaters~\citet{Briegel1998,Sangouard2011}
promise to overcome this problem. The idea is to split the distance
between the network parties into elementary segments comparable with
the fiber attenuation length, such that the high-fidelity entanglement
in the segments could be distributed independently. The entangled
states of the segments are generated probabilistically and are kept
in quantum memories while others segments are under preparation. After
that, the length of the entangled segments is extended using the quantum
swapping operations~\citet{Bennett1993} in a nested way~\citet{Duan2001}.

Recent progress in quantum hardware gave rise to experimental implementation
of individual components of the quantum repeater networks bringing
closer realization of the first large-scale quantum internet. A non-exhaustive
list of the experimental achievements includes: demonstration of generation
of a distributed pair of entangled qubits~\citet{Liao2018,Humphreys2018,Wengerowsky2019};
long-coherence time, high efficiency and mode-multiplexing for quantum
memories are separately presented in~\citet{Zhong2015,Kim2016,Wang,Pu2017,Zhong2017,Laplane2017,Katz2018,Dou2017,Korber2017,Bradley2019,Wang2019};
interfacing of atomic memories with telecom fiber~\citet{Seri2017,Dreau2018,Rancic2018,Askarani2019,Krutyanskiy2019,VanLeent2019};
swapping operation for qubits pairs~\citet{Su2016,Sun2017,Tsujimoto2018,Zopf2019}.
The experimental progress leads to a theoretical challenge to analyze
and optimize properties of real-world large-scale quantum networks
taking into account all essential experimental imperfections.

The repeater networks subjected to such hardware imperfections as
detectors inefficiency and dark counts, losses in fibers, classical
communication delays, quantum memory read-out inefficiency, and memory
decoherence were considered in a number of papers. In particular,
the numerical and analytical study of quantum repeaters where presented
in~\citet{Collins2007,Brask2008,Sangouard2011,Shchukin2017,Santra2019}.
These works, however, considered simplified models for the quantum
memory decoherence. For instance, in Ref.~\citet{Collins2007} the
quantum memory efficiency is considered perfect until some finite
coherence time $t_{\text{coh}}$ after which the information is lost
entirely. Authors of~\citet{Brask2008,Sangouard2011} study effects
of finite memory read-out efficiency in the limit of infinitely large
memory coherence time. In Ref~\citet{Santra2019}, a pure dephasing
of memories and classical communication time in a one-dimensional
(1D) repeater protocol were considered without other experimental
imperfections.

The difficulty of modeling the protocols of quantum repeaters with
a general time-continuous memory decoherence originates in the stochastic
evolution of the networks. This evolution consists of random preparation
of the elementary segments followed by probabilistic operations of
entanglement swapping \citet{Duan2001,Borregaard2015}. A possible
solution for exact simulation of such a probabilistic network is the
Monte Carlo (MC) method, as was presented in Ref.~\citet{Zippilli2016,Matsuo2018,Dahlberg2019,Kuzmin2019}.
Using this method one can recover statistics of a quantum network
by simulating probabilistic trajectories, each of which represents
execution of the full repeater protocol step by step. While being
very flexible, the MC method has a limiting drawback — its runtime
is proportional to the entanglement generation time in the simulated
quantum network, which is quickly increasing with the network size.
Therefore, the MC method becomes impractical for the performance evaluation
and optimization of the real-world large-scale networks.

The purpose of the present paper is to elaborate on the alternative,
semi-analytical, diagrammatic technique, which can be used to efficiently
analyze the realistic quantum repeaters networks, as was demonstrated
in our previous work~\citet{Kuzmin2019}. This technique takes into
account all essential experimental imperfections such as fiber losses,
detectors losses, dark counts in detectors, memory read-out inefficiency,
classical communication time, and the effect of dissipative Liouville
dynamics on the network quantum memories (decay, dephasing, etc).
The semi-analytical technique requires several orders of magnitude
less computational run-time than the MC method while keeping the high
precision of the simulation, as shown below by direct comparison of
the simulations results.

The paper is organized as follows. In Sec.~\ref{sec:Methods-and-results}
we introduce the repeater setup and explain the basic idea of the
semi-analytic technique. Then we move on to develop the diagrammatic
method step by step. In Sec.~\ref{sec:Elementary-segment-generation}
we introduce the elementary diagrams used in the semi-analytical method.
In Sec.~\ref{sec:Memories-decay-and} we derive basic equations for
the 1D and 2D protocols which take into account imperfections in generation
of the elementary repeater segments, imperfect merging operations,
and the effect of dissipation on the network quantum memories. In
Sec.~\ref{sec:Communication-time} we show how the diagrammatic technique
incorporates the time for classical communications between the network
nodes. In Sec.~\ref{sec:Temporal-filtering-protocol} we develop
diagrams for describing the temporal filtering protocol~\citet{Santra2019,Kuzmin2019},
which improves the fidelity of the states generated by the repeater
networks in presence of finite memories time. In Sec.~\ref{sec:Benchmarking-with-Monte}
we benchmark the accuracy of the developed method against the MC simulation.
Finally, in Sec.~\ref{sec:Network-optimization}, we demonstrate
a real-world application of the method — optimization of the network
parameters for maximization of the secret key generation rate, and
we conclude in Sec.~\ref{sec:Outlook}.

\section{The quantum network setup and the semi-analytic method\label{sec:Methods-and-results}}

\begin{figure}
\includegraphics{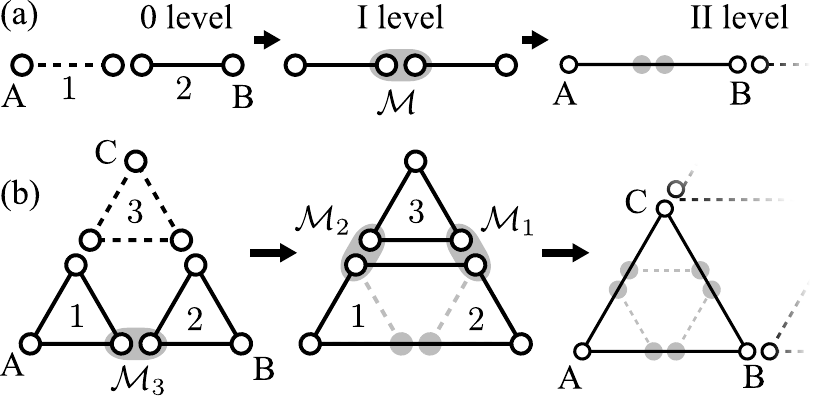}\caption{Nesting level schemes of probabilistic quantum repeaters: (a) 1D repeater
for the distribution of the Bell state $\left(\left\vert 01\right\rangle +\left\vert 10\right\rangle \right)/\sqrt{2}$
and (b) 2D repeater distributing the GHZ state $\left(\left\vert 110\right\rangle +\left\vert 001\right\rangle \right)/\sqrt{2}$.
Circles represent quantum memories. Solid~(dashed) lines indicate
(dis-)\hspace*{0cm}entanglement of the connected memories. $\mathcal{M}$
and $\mathcal{M}_{i}$ are the superoperators representing merging
operations applied to the pairs of quantum memories within the gray
shaded areas. The gray colored quantum memories are traced out of
the segments states.}
\label{fig:1}
\end{figure}

We start our discussion with a summary of the approach developed in~\citet{Kuzmin2019}
for analysis of the performance of large scale quantum repeater networks.
A quantum network is a nested structure consisting of entangled nodes
represented by quantum memories. Examples of one and two dimensional
quantum networks are shown in the Fig.~\ref{fig:1}. The quantum
repeater idea is to generated entanglement between the most remote
nodes (circles in Fig.~\ref{fig:1}) of the network by probabilistic
generation of entanglement between the adjacent nodes followed by
a sequence of entanglement swapping operations. This allows to grow
the size of the entangled segments (nodes connected by solid lines
in Fig.~\ref{fig:1}) of the network until the most remote nodes
become entangled.

On a more detailed level the quantum repeater protocol is described
as follows. First, each elementary segment $i$, in the nesting level
$0$ of a network, is probabilistically prepared in an entangled state
described by the density matrix $\rho_{i}$. The state is stored in
quantum memories subjected to decoherence. Next, the neighboring segments
of the network are probabilistically merged by applying the swapping
operation to quantum memories within one node. Upon success, a larger
number of nodes become entangled, thus, forming a network segment
of the next nesting level. If the merging fails, the two segments
has to be prepared again from scratch. The process is executed recursively
until the final network state is prepared. 

\begin{figure}
\includegraphics{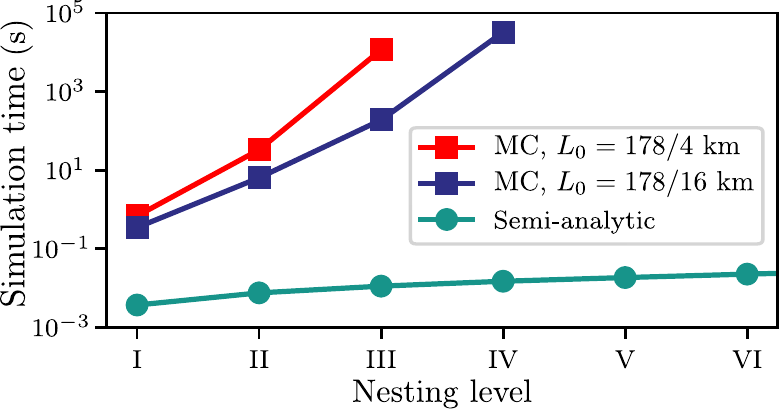} \caption{Computation run time versus the network nesting level $N$. The physical
distance covered by the network grows as $2^{N}L_{0}$ with $L_{0}$
the length of the elementary segment. Square points show the MC simulation
run time required to achieve $1\%$ standard deviation for the fidelity
values. Red and blue colors show the two network sizes simulated by
MC method: $L_{0}=178/16$ km and $L_{0}=178/4$ km. Round points
show the computation time with the semi-analytical method, which does
not depend on $L_{0}$.}
\label{fig:-2}
\end{figure}

The challenge is to estimate performance of such a network in the
presence of realistic imperfections such as photon loss, communication
delay, quantum memory decay, etc. Here, we address the problem using
the semi-analytic method devised in~\citet{Kuzmin2019}. In this
approach, unlike in the MC method, we derive analytical equations
describing statistics of a single nesting level of the network. Thanks
to the nested structure of the repeater protocol we recursively reuse
the derived equations to evolve states of the elementary segments
up to the final network level. 

The main idea is to determine the density matrix distribution $\varrho(t)$
for the ensemble of states generated by the network at time $t$.
More precisely, $\varrho(t)dt$ is the unnormalized density matrix
of the ensemble of states generated within the time window $[t,t+dt)$,
and $\text{Tr}\left\{ \varrho(t)dt\right\} $ is the probability to
generate the state within this time window. In the following we show
how to obtain the Laplace image of the distribution, $\tilde{\varrho}(s)=\int_{0}^{\infty}e^{-st}\varrho(t)\,dt$,
which fully describes the statistics of the network and allows one
to infer the average generated state $\rho$ and corresponding generation
time $T$:

\begin{gather}
\begin{aligned}\rho & =\int_{0}^{\infty}\varrho(t)dt=\tilde{\varrho}(s)\Big|_{s=0},\\
T & =\int_{0}^{\infty}t\,{\rm Tr}\left\{ \varrho(t)\right\} dt=-{\rm Tr}\left\{ \frac{d\tilde{\varrho}(s)}{ds}\right\} \Big|_{s=0}.
\end{aligned}
\label{eq:-44}
\end{gather}

The method for deriving the equation for the density matrix distribution
$\varrho(t)$ allows us to include a general time-continuous decoherence
process which can be described by a Master Equation $\dot{\rho}_{i}=\mathcal{L}_{i}\rho_{i}$
with the Lindblad superoperator $\mathcal{L}_{i}$. The most time
consuming part of the numerical calculation stage of the semi-analytical
approach is solving a sparse system of linear equations involving
the superoperator $\mathcal{L}_{i}$. This is a standard numerical
problem which can be solved very efficiently. As a result we have
a very fast method for simulation of quantum networks as shown in
Fig.~\ref{fig:-2}~\footnote{The plot shows the computation time using one core of Apple MacBook
Air with a 22-nm 1.7 GHz Intel Core i7 processor (4650U).}. The computation run-time grows only linearly with the number of
nesting levels in the network and it is independent of the values
of the network parameters. In contrast, the run-time for the MC simulation
(implemented according to~\citet{Kuzmin2019}) grows faster than
exponential and depends on the elementary segment size $L_{0}$, limiting
the size of the simulated networks. Below, in Sec~\ref{sec:Benchmarking-with-Monte},
we demonstrate that the results obtained by our approach match the
MC simulation results.

The semi-analytical method involves summation of infinite series corresponding
to averaging over all possible trajectories leading to generation
of the final network state. Therefore, in the following Sections we
introduce diagrams to facilitate the evaluation of the infinite sums
representing density matrix distributions for quantum networks of
high complexity.

\begin{comment}
In the following we outline the diagrammatic technique using 1D network
as an example. We consider two segments~(links) of the 1D network,
shown in Fig.~\ref{fig:1}(a), with states $\rho_{1}$ and $\rho_{2}$
which are merged into a longer segment with an average state $\rho$
to be found. We assume that a probability $q$ to generate an elementary
segment $i$ in a single attempt is small, $q\ll1$~\citet{Sangouard2009},
and requires time $\Delta t$. Therefore, we can introduce a continuous
probability density $p(t)=\nu\text{\,exp}\left(-\nu t\right)$, with
rate $\nu=q/\Delta t$, describing generation of elementary segments.
The Laplace image of the distribution can be easily obtained as $\tilde{p}(s)=\int_{0}^{\infty}e^{-st}p(t)\,dt=\nu/(\nu+s)$. 
\end{comment}

\begin{comment}
Now, using the probability distribution, we can introduce the \emph{density
matrix} distribution for generation of the $i^{\text{th}}$ link in
the state $\rho_{i}$ at time $t$:

Then, the probability density for generating the $i^{\text{th}}$
link before the second, $j^{\text{th}}$ link has been generated at
time $t$ is $p(t)\int_{0}^{t}dt'\,p(t')$ with the corresponding
Laplace image $\nu^{2}/[(s'+\nu)s']\mid_{s'=s+\nu}$.
\end{comment}
{} %
\begin{comment}
Here $\left(ij\right)\in\{\left(12\right),\left(21\right)\}$ defines
the random generation order, and we integrate over all possible times
$t'$ of the $i^{\text{th}}$ link creation.
\end{comment}
{} %

\section{Diagrammatic technique\label{subsec:1D-repeater}}

In the section, we develop the diagrammatic technique for evaluation
of the density matrix distribution describing merging of the network
segments and, consequently, the full repeater protocol. First, we
consider generation of the elementary segments, and then proceed with
the description of the bipartite and tripartite networks based on
the 1D and 2D repeater protocols, correspondingly. %
\begin{comment}
Finite communication time and the temporal filtering protocol are
omitted in this section and are introduces in Sec.~\ref{sec:Communication-time}
and Sec.~\ref{sec:Temporal-filtering-protocol} correspondingly.
\end{comment}

\subsection{Generation of elementary segments \label{sec:Elementary-segment-generation}}

First, we introduce diagrams representing generation of the elementary
repeater segments (links in 1D case), which are the building blocks
in the diagrammatic technique. Let us consider an elementary segment
of a repeater network generated in discrete steps with success probability
$q$ for each attempt. We assume $q\ll1$ (typical for quantum repeaters~\citet{Duan2001})
what results in a large number of generation attempts $K\gg1$. Then,
the probability of generating the elementary segment in $K$ attempts
is
\begin{align}
q_{K} & =q\left(1-q\right)^{K-1}\label{eq:13}\\
 & =qe^{\left(K-1\right)\text{ln}\left(1-q\right)}\approx qe^{-Kq}.\nonumber 
\end{align}

Defining the time for one generation step as $\Delta t$, one can
introduce an average link generation rate $\nu=q/\Delta t$, total
link generation time $t=K\Delta t$, and the continuous probability
distribution 
\begin{align}
p(t) & =\nu e^{-\nu t},\label{eq:-31}
\end{align}
with the Laplace image
\begin{equation}
\tilde{p}(s)=\frac{\nu}{s+\nu}.\label{eq:-43}
\end{equation}

Now, we introduce the elementary diagram representing the density
matrix distribution for generation of a segment $i$ in the state
$\rho_{i}$ at time $t$:
\begin{equation}
\varrho_{i}(t)=p(t)\rho_{i}=\nu e^{-\nu t}\rho_{i}\equiv\diagram{diag1}.\label{d1}
\end{equation}

\noindent Here the vertical line of length $t$ represents the total
link generation time, and the circle denotes the successful generation
event. Time flows upward.

\subsection{1D repeater\label{sec:Memories-decay-and}}

Consider a 1D repeater network consisting of two links with states
$\rho_{1}$ and $\rho_{2}$ generated at the rate $\nu$, as shown
in Fig.~\ref{fig:1}(a). The goal is to find the average state $\rho$
and the generation time $T$ of the segment in the next nesting level
obtained by merging the two links. After that, we can define a recursive
procedure to find average states generated at all nesting levels of
the network.

\paragraph*{Generation of segments within one nesting level.}

We proceed starting with the density matrix distribution for generating
the $i^{\text{th}}$ link in state $\rho_{i}$ before the $j^{\text{th}}$
link has been generated at time $t$ in state $\rho_{j}$ which can
be constructed using two diagrams $\eqref{d1}$,

\begin{gather}
\varrho_{ij}(t)=\diagram{322}\nonumber \\
=p(t)\int_{0}^{t}dt'\,p(t')e^{(t-t')\mathcal{L}{}_{i}}\rho_{\text{in}}\equiv X_{ij}\left(t\right)\rho_{\text{in}},\label{eq:-19}
\end{gather}

\noindent with $\rho_{\text{in}}=\rho_{1}\otimes\rho_{2}$. Here we
introduce propagator $X_{ij}\left(t\right)$ which includes integration
over all intermediate times $t'$ of the $i^{\text{th}}$ link creation.
The degradation of the $i^{\text{th}}$ link during the waiting time
$t-t'$ due to the finite lifetime of the quantum memory is represented
by the dashed line and is taken into account using the Lindblad superoperator
$\mathcal{L}_{i}$~\footnote{The decay of the corresponding quantum memories is described by the
Master Equation $\dot{\rho}_{i}=\mathcal{L}_{i}\rho_{i}$ with the
Lindblad superoperator $\mathcal{L}_{i}$:
\[
\mathcal{L}_{i}\bullet=T_{\text{coh}}^{-1}\sum_{k}\left[c_{k}^{(i)}\bullet c_{k}^{(i)\dagger}-\frac{1}{2}\{c_{k}^{(i)\dagger}c_{k}^{(i)},\bullet\}_{+}\right],
\]
where $c_{k}^{(i)}$ is the jump operator describing a desired model
of decoherence for the $k^{{\rm th}}$ quantum memory of the segment
$i$, the black dot is a placeholder for the density matrix, and $\{\ldots\}_{+}$
corresponds to the anticommutator.}. Inserting the probability distributions $\eqref{eq:-31}$ into Eq.~$\eqref{eq:-19}$
we find the Laplace image of the density matrix distribution $\tilde{\varrho}_{ij}(s)=\tilde{X}_{ij}(s)\rho_{\text{in}}$
with

\noindent 
\begin{align}
\tilde{X}_{ij}(s) & =\int_{0}^{\infty}dt\,e^{-st}X_{ij}\left(t\right)\nonumber \\
 & =\nu^{2}\int_{0}^{\infty}dt\,e^{-\left(s+\nu\right)t}\int_{0}^{t}dt'\,e^{-\nu t'}e^{(t-t')\mathcal{L}{}_{i}}\nonumber \\
 & =\frac{\nu^{2}}{(s'+\nu)(s'-\mathcal{L}{}_{i})}\big|_{s'=s+\nu}.\label{eq:-32}
\end{align}
Here we use the fact that the convolution of two functions in Eq.~\eqref{eq:-19}
is transformed into a product of their Laplace images and the exponential
prefactor shifts the frequency of the Laplace images.

Summing over two possible orders of the segments generation $(i,j)=\{(1,2),(2,1)\}$
we obtain the density matrix distribution and its Laplace image for
the two segments prepared for merging at time $t$,
\begin{align}
\varrho_{\text{prep}}\left(t\right) & =\sum_{ij}\varrho_{ij}(t)=\diagram{3}+\diagram{2},\label{d2}\\
\tilde{\varrho}_{\text{prep}}\left(s\right) & =\sum_{ij}\tilde{\varrho}_{ij}(s)=\sum_{ij}\tilde{X}_{ij}(s)\rho_{\text{in}}\nonumber \\
 & =\frac{\nu^{2}}{(s'+\nu)}\left[\frac{1}{(s'-\mathcal{L}{}_{1})}+\frac{1}{(s'-\mathcal{L}{}_{2})}\right]\Big|_{s'=s+\nu}\rho_{\text{in}}.\label{eq:rho_prep(s)_1D}
\end{align}

\paragraph*{Merging of the generated segments.}

Once both segments are prepared, we perform a probabilistic merging
operation described by a non-trace-preserving superoperator $\mathcal{M}$~(see
Appendix~\ref{Appendix: Calculation-of-the}). A successful merging
of the two prepared segments with \emph{one} attempt at time $t$
results in the longer segment with the density matrix distribution
given by
\begin{align}
 & \mathcal{M}\varrho_{\text{prep}}(t)\equiv\diagram{4}+\diagram{5}\equiv\diagram{6},\label{d3}
\end{align}
here we introduce a new diagram to represent the merging result. The
probability density for a failed merging operation at time $t$ reads
\begin{align}
 & \text{Tr}\left\{ \left(\mathcal{I}-\mathcal{M}\right)\varrho_{\text{prep}}\left(t\right)\right\} \equiv\diagram{7},\label{d4}
\end{align}

\noindent where $\mathcal{I}$ is the unit superoperator. Next, a
diagram describing a successful merging at time $t$ with one failed
attempt at time $t_{0}$ is constructed out of diagrams~\eqref{d3}
and~\eqref{d4} as,

\begin{gather}
\diagram{8}\nonumber \\
=\int_{0}^{t}dt_{0}\ \mathcal{M}\varrho_{\text{prep}}(t)\cdot\text{Tr}\left\{ \left(\mathcal{I}-\mathcal{M}\right)\varrho_{\text{prep}}\left(t_{0}\right)\right\} ,\label{eq:-51}
\end{gather}

\noindent where we integrate over all possible times of the failed
attempt $0<t_{0}<t$. Distribution~\eqref{eq:-51} is a convolution
and its Laplace image is a product of the Laplace images of Eqs.~\eqref{d3}
and \eqref{d4}, $\mathcal{M}\tilde{\varrho}_{\text{prep}}(s)\cdot\text{Tr}\left\{ \left(\mathcal{I}-\mathcal{M}\right)\tilde{\varrho}_{\text{prep}}(s)\right\} $.

\paragraph*{Sum over all trajectories.}

The target density matrix distribution $\varrho(t)$ of the longer
segment is obtained by summing over all possible combinations of unsuccessful
mergings leading to a final successful generation of an entangled
state 
\begin{gather}
\varrho(t)=\diagram{6}+\diagram{8}+\diagram{9}+\dots\label{eq:-2}
\end{gather}
In the Laplace domain, the sum~\eqref{eq:-2} corresponds to a geometric
series, which converges to

\begin{align}
\tilde{\varrho}(s) & =\mathcal{M}\tilde{\varrho}_{\text{prep}}\left(s\right)\sum_{m=0}^{\infty}\big[\text{Tr}\left\{ \left(\mathcal{I}-\mathcal{M}\right)\tilde{\varrho}_{\text{prep}}\left(s\right)\right\} \big]^{m}\nonumber \\
 & =\frac{\mathcal{M}\tilde{\varrho}_{\text{prep}}\left(s\right)}{1-\text{Tr}\left\{ \left(\mathcal{I}-\mathcal{M}\right)\tilde{\varrho}_{\text{prep}}\left(s\right)\right\} }.\label{eq:-47}
\end{align}

Finally, inserting Eq.~\eqref{eq:rho_prep(s)_1D} into Eq.~$\eqref{eq:-47}$
and taking into account that $\text{Tr}\left\{ {\cal L}_{i}\rho_{i}\right\} =0$,
we use Eq.~$\eqref{eq:-44}$ to obtain the average state and generation
time

\noindent 
\begin{align}
\begin{aligned}\rho & =\tilde{\varrho}(s)\Big|_{s=0}=\frac{\mathcal{M}\tilde{\varrho}_{\text{prep}}\left(0\right)}{\text{Tr}\left\{ \mathcal{M}\tilde{\varrho}_{\text{prep}}\left(0\right)\right\} }\\
 & =\frac{1}{P}\cdot\frac{1}{2}\mathcal{M}\left(\frac{1}{1-{\cal L}_{1}/\nu}+\frac{1}{1-{\cal L}_{2}/\nu}\right)\rho_{1}\otimes\rho_{2}\\
T & =-{\rm Tr}\left\{ \frac{d\tilde{\varrho}(s)}{ds}\right\} \Big|_{s=0}=\frac{1}{P}\cdot\frac{3}{2\nu},
\end{aligned}
\label{eq:-1}
\end{align}

\noindent 

\noindent with $P=\text{Tr}\left\{ \mathcal{M}\tilde{\varrho}_{\text{prep}}\left(0\right)\right\} $
the average merging probability. One can see that the probability
of the eventual state preparation $\text{Tr}\left\{ \rho\right\} =1$.
This indicates that we take into account all possible trajectories.
The resulting Eqs.~$\eqref{eq:-1}$ allow us to find the state and
the generation time of the next nesting level segment using efficient
numerical algorithms for solving the sparse systems of linear equations
involving ${\cal L}_{i}$.

\paragraph*{Full repeater protocol.}

To address the next nesting level of the network, we assume that its
segments are generated time-independently in states \textbf{$\rho'_{i}\equiv\rho$}
with rates $\nu'\equiv1/T$ obtained from Eqs.~$\eqref{eq:-1}$.
In other words, we assume that the generation of the segments in the
next network levels is a Poissonian process, and Eqs.~$\eqref{eq:rho_prep(s)_1D}$
and $\eqref{eq:-1}$ can be reused to evaluate the properties of the
higher nesting levels segments. This is an approximation since the
exact density matrix distribution $\varrho(t)$ given by its Laplace
image~\eqref{eq:-47} describes preparation of states which are time-dependent.
This is, however, a very good approximation as we show below. 

Using the described procedure recursively, we can find states and
generation times of all nesting levels of the network. Notice, that
Eqs.~$\eqref{eq:rho_prep(s)_1D}$ and $\eqref{eq:-1}$ operate with
a quantum state of at most 4 quantum memories, therefore, the calculation
of the desired quantities does not demand a lot of computational resources
in comparison with the MC method.

Finally, we mention that to start the described recursion, one has
to obtain states $\{\rho_{i}\}$ and generation rate $\nu=q/\Delta t$
of the elementary segments, and compute the merging superoperator
$\mathcal{M}$. To obtain $\{\rho_{i}\}$, $q$ and $\Delta t$ one
can simulate the protocol for preparation of the elementary segment,
for example, in terms of density matrices taking into account all
essential experimental imperfections. Superoperator $\mathcal{M}$
is calculated based on the merging protocol taking into account the
imperfections as well. An example of simulation of the elementary
segments and construction of $\mathcal{M}$ for the DLCZ protocol
is presented in Appendix~\ref{Appendix: Calculation-of-the}.

\paragraph*{Verification of the approximations validity.}

In the following, we illustrate the validity of the approximation
of regarding the generation of the segments in the next network levels
as a Poissonian process. We consider the case of no dissipation, $\mathcal{L}{}_{i}=0$,
and compare the exact generation probability distribution $r(t)=\text{Tr}\left\{ \varrho(t)\right\} $
for the segment generated in the first nesting level with the Poissonian
distribution of rate $1/T$ defined in Eqs.~\eqref{eq:-1}.

\begin{figure}
\includegraphics{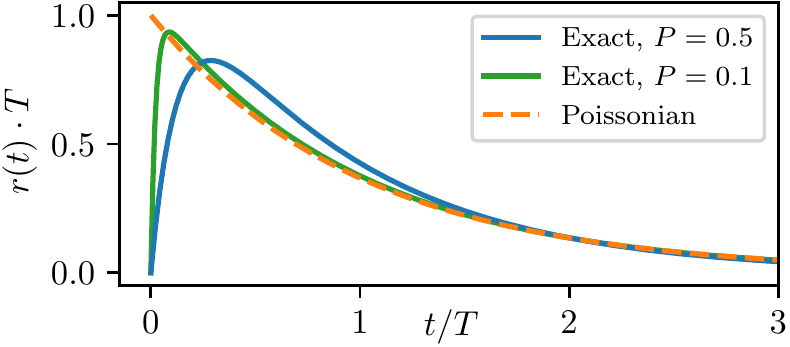} \caption{Exact distributions to generate a state of the first nesting level
for various probabilities for the merging success $P$ and their approximation
by the Poissonian distribution. $T$ is the average generation time
from Eqs.~\eqref{eq:-1}.}
\label{fig:-4}
\end{figure}

First, we obtain the Laplace image of $r(t)$ as $\tilde{r}(s)=\text{Tr}\left\{ \tilde{\varrho}\left(s\right)\right\} $.
From Eq.~\eqref{eq:rho_prep(s)_1D}, in the case of $\mathcal{L}{}_{i}=0$,
we obtain
\begin{align*}
\text{Tr}\{\mathcal{M}\tilde{\varrho}_{\text{prep}}\left(s\right)\} & =P\cdot\text{Tr}\{\tilde{\varrho}_{\text{prep}}\left(s\right)\}\\
\text{Tr}\{\tilde{\varrho}_{\text{prep}}\left(s\right)\} & =\frac{2\nu^{2}}{(s+2\nu)(s+\nu)}
\end{align*}
Then, using the above equations and Eq.~\eqref{eq:-47}, one can
obtain

\begin{equation}
\tilde{r}(s)=\frac{2P\nu^{2}}{s^{2}+3\nu s+2P\nu^{2}}.\label{eq:-23}
\end{equation}
This function has two poles in the left half-plane at points $3\nu(-1\pm\sqrt{1-(8/9)P})/2$,
each of which, in the time domain, corresponds to an additive exponential
term with the decay rate given by the pole. In the limit of small
merging success probability, $P\ll1$, the first-order Tailor expansion
reveals that one pole $a\approx-3\nu+(2/3)P\nu$ gives an exponent
decaying much faster than the exponent arising from the second pole
$b\approx-(2/3)P\nu$. As a result, we obtain the probability distribution
as

\begin{align*}
r(t) & =2P\nu^{2}\frac{1}{a-b}\left(e^{at}-e^{bt}\right)\\
 & \underset{P\ll1,\,\nu t\gg1}{\longrightarrow}\frac{2}{3}P\nu e^{-\frac{2}{3}P\nu t}=\frac{1}{T}e^{-\frac{t}{T}}
\end{align*}
with the asymptotic behavior approaching the Poissonian distribution
with rate $1/T$, where $T$ is defined by Eqs.~\eqref{eq:-1}. This
behavior is demonstrated in Fig.~\ref{fig:-4} for $P=0.5$ and $P=0.1$.
For realistic networks this approximation is fulfilled well since
the merging probability is at most $0.5$ for an ideal setup and drops
quickly with the addition of realistic imperfections and increase
of the nesting level of the network.

\subsection{2D repeater\label{subsec:2D-repeater}}

In the subsection, we develop the diagrammatic technique for the protocol
of the 2D repeater illustrated in Fig.~$\ref{fig:1}$(b). The 2D
repeater is implemented in a nested way, such that, at each nesting
level, segments $i=\{1,2,3\}$ with states $\rho_{i}$ and generation
rates $\nu$ are probabilistically merged to form the segment of the
next nesting level with the state $\rho$. The segments are generated
in random order $ijk$, which is one of the possible permutations
of the segments indices $\{123\}$. Merging of the first pair of prepared
segments, $i$ and $j$, is described by the superoperator $\mathcal{M}_{k}$
(see Appendix~\ref{Appendix: Calculation-of-the}). Merging of the
resulting four-node state with the third prepared segment $k$ is
implemented with two merging operations $\mathcal{M}_{i}$ and $\mathcal{M}_{j}$,
which can be combined in one superoperator $\bar{\mathcal{M}}_{k}\equiv\mathcal{M}_{i}\mathcal{M}_{j}$.

Now we show how to find the density matrix distributions for the 2D
repeater protocol reusing some of the results obtained for the 1D
repeater in the previous section. First, let us consider the following
equality for diagrams describing a set of trajectories for 3 segments
of the 2D repeater including one successful and one unsuccessful attempts
to merge two segments
\[
\diagram{602}+\diagram{601}+\diagram{600}=\diagram{603}.
\]
The diagrams in the left part of the equality illustrate all 6 possible
variants for a failed merging of a pair of segments at time $t_{0}$
followed by a successful merging of a pair of segments (prepared in
two possible orders). Then, the successfully merged at time $t_{1}$
state degrades until the third segment is generated at time $t$.
In the diagrams, each vertical line of some length $t'$ represents
probability that a segment, being generated with the rate $\nu$,
is not prepared at the time $t'$ and, in time domain, gives a prefactor
$\text{exp}(-\nu t')$ to the density matrix distribution. Therefore,
two such vertical lines of length $t'_{1}$ and $t'_{2}$ can be combined
in a single line of length $t'_{1}+t'_{2}$. Thus, in the left part
of the equality, we can reshuffle parts of the diagrams to collect
the blue colored lines a single column. This can be done if the segments
have identical generation rates $\nu$. Afterward, the parts of the
diagrams corresponding to the failed mergings can be factored out
and summed to a new diagram which we illustrated with bold lines.
In the following we omit the blue color of the line.

Now we can represent all possible trajectories for merging of two
segments $i,\,j\ne k$ at time $t$ as

\begin{gather}
\diagram{205}+\diagram{604}+\diagram{605}+\dots\equiv\diagram{606}\label{eq:-18}\\
=p(t)\int_{0}^{t}dt_{0}\,e^{\overline{\mathcal{L}}_{k}(t-t_{0})}Y_{k}(t_{0})\rho_{\text{in}}.\label{eq:-36}
\end{gather}
with $\rho_{\text{in}}=\rho_{1}\otimes\rho_{2}\otimes\rho_{3}$. Here
$p(t)$ given by Eq.~$\eqref{d1}$ is the probability distribution
for generation of the last prepared segment $k$. Propagator $Y_{k}(t)$
generates a merged state of segment $i,\,j\ne k$ at time $t_{0}$
averaged over all possible trajectories, and superoperator $\overline{\mathcal{L}}_{k}$
describes degradation of segments $i$ and $j$ after their merging.
The Laplace image of Eq.~\eqref{eq:-36} is
\begin{equation}
\frac{\nu}{s'-\overline{\mathcal{L}}_{k}}\tilde{Y}_{k}(s')\Big|_{s'=s+\nu}\rho_{\text{in}}.\label{eq:-21}
\end{equation}
Here $\tilde{Y}_{k}(s')$ is the Laplace image of $Y_{k}(t)$, which
can be obtained in analogy to Eq.~\eqref{eq:-47} by summing a geometric
series of diagrams with all possible variants of the failed mergings

\begin{multline*}
\tilde{Y}_{k}(s)=\frac{\mathcal{M}_{k}}{1-\tilde{p}_{\text{fail}}^{(1)}(s)}\left[\tilde{X}_{ij}(s)+\tilde{X}_{ji}(s)\right],\:i,j\ne k,
\end{multline*}
In the above equation, $\tilde{X}_{ij}(s)$ and $\tilde{X}_{ji}(s)$
given by Eq.~\eqref{eq:-32} describe two orders of generation of
the merged segments $i$ and $j$. At the same time, the Laplace image
$\tilde{p}_{\text{fail}}^{(1)}(s)$ of the probability distribution
for a failed merging includes all 6 possible generation orders \{$ijk$\}
which can lead to a failed merging

\begin{multline*}
\tilde{p}_{\text{fail}}^{(1)}(s)=\text{Tr}\Big\{\big([\mathcal{I}-\mathcal{M}_{1}][\tilde{X}_{23}(s)+\tilde{X}_{32}(s)]\\
+[\mathcal{I}-\mathcal{M}_{2}][\tilde{X}_{13}(s)+\tilde{X}_{31}(s)]\\
+[\mathcal{I}-\mathcal{M}_{3}][\tilde{X}_{12}(s)+\tilde{X}_{21}(s)]\big)\rho_{\text{in}}\Big\}.
\end{multline*}

The diagram for a successful attempt of the third, last, merging is
represented as
\[
\sigma(t)=\diagram{607}
\]
and includes three possible variants with one of three segments $k\in\{1,2,3\}$
being prepared at the end. In analogy with Eq.~\eqref{d3}, we use
Laplace image~\eqref{eq:-21} for diagram~\eqref{eq:-18} to find
the Laplace image of $\sigma(t)$ as
\[
\tilde{\sigma}(s)=\sum_{k=\{1,2,3\}}\bar{\mathcal{M}}_{k}\frac{\nu}{s'-\overline{\mathcal{L}}_{k}}\tilde{Y}_{k}(s')\Big|_{s'=s+\nu}\rho_{\text{in}}.
\]

The diagram for the corresponding failed merging is
\[
p_{\text{fail}}^{(2)}(t)=\diagram{612}
\]
with the Laplace image obtained by replacing $\bar{\mathcal{M}}_{k}$
with $I-\bar{\mathcal{M}}_{k}$ in the equation for $\tilde{\sigma}(s)$
and taking its trace

\begin{multline*}
\tilde{p}_{\text{fail}}^{(2)}(s)=\sum_{k=\{1,2,3\}}\text{Tr}\Bigg\{\left(I-\bar{\mathcal{M}}_{k}\right)\frac{\nu}{s'-\overline{\mathcal{L}}_{k}}\\
\times\tilde{Y}_{k}(s')\Big|_{s'=s+\nu}\rho_{\text{in}}\Bigg\}.
\end{multline*}

Then, similar to diagrams series~\eqref{eq:-2}, the distribution
for the final state preparation can be written as an infinite sum

\begin{gather}
\varrho(t)=\diagram{607}+\diagram{608}+\diagram{609}+\dots.\label{eq:-22}
\end{gather}

The diagrams in Eq.~\eqref{eq:-22} becomes a geometric series in
the Laplace domain and its sum equals to
\begin{equation}
\tilde{\varrho}(s)=\frac{\tilde{\sigma}(s)}{1-p_{\text{fail}}^{(2)}(s)}.\label{eq:-33}
\end{equation}
Introducing the average success probability for the first merging
as $P_{1}=1-\tilde{p}_{\text{fail}}^{(1)}(0)$ and for the combination
of the second and the third mergings as $P_{2}=1-\tilde{p}_{\text{fail}}^{(2)}(0)$
one can find using Eqs.~\eqref{eq:-44} the average state and generation
time as

\noindent 
\begin{gather}
\begin{aligned}\rho & =\tilde{\varrho}(s)\Big|_{s=0}=\frac{1}{6P_{1}P_{2}}\sum_{k=\{1,2,3\}}\bar{\mathcal{M}}_{k}\frac{1}{1-\overline{\mathcal{L}}_{k}/\nu}\\
 & \times\mathcal{M}_{k}\sum_{i\ne k}\frac{1}{1-\mathcal{L}{}_{i}/\left(2\nu\right)}\rho_{\text{in}},\\
T & =-{\rm Tr}\left\{ \frac{d\tilde{\varrho}(s)}{ds}\right\} \Big|_{s=0}=\frac{5}{6P_{1}P_{2}\nu}.
\end{aligned}
\label{eq:-24}
\end{gather}

In analogy with the 1D repeater, here we demonstrate that the next-nesting-level
network can be calculated using the same Eq.~\eqref{eq:-24} with
the input given by the average density matrix $\rho$ and the average
generation rate $1/T$ obtained for the current nesting level. To
do that, we consider probability distribution $r(t)$ for generation
of a segment in the first nesting level in the case with no decoherence,
$\mathcal{L}_{k}=0$ and $\overline{\mathcal{L}}_{k}=0$ and compair
it with the Poissonian distribution. Under this simplification, the
Laplace image of the distribution $\tilde{r}(s)=\text{Tr}\{\tilde{\varrho}(s)\}$
can be obtained by replacing in the above equations all the merging
superoperators to the average merging probabilities, $\mathcal{M}_{k}\to P_{1}$
and $\bar{\mathcal{M}}_{k}\to P_{2}$, and then taking the trace.
As a result, one can find

\begin{equation}
\tilde{r}(s)=\frac{6P_{1}P_{2}\nu^{3}}{s^{3}+6s^{2}\nu+(5+6P_{1})s\nu^{2}+6P_{1}P_{2}\nu^{3}}.\label{eq:-37}
\end{equation}
This function has three poles, which in the limit of small merging
success probabilities, $P_{1},P_{2}\ll1$, have the dominating parts
$-6/5P_{1}P_{2}\nu$, $-5\nu$, and $-\nu$. In the time domain, the
first pole results in an exponent which has the slowest decay rate
and, therefore, gives the asymptotic behavior of the probability distribution
\[
r(t)\underset{P_{1},P_{2}\ll1,\,\nu t\gg1}{\to}\frac{6}{5}P_{1}P_{2}\nu e^{-\frac{6}{5}P_{1}P_{2}\nu t}=\frac{1}{T}e^{-\frac{t}{T}}
\]
which approaches the Poissonian distribution with rate $1/T$ for
$t\nu\gg1$ with $T$ given by Eqs.~\eqref{eq:-24}. The validity
of this approximation for the case with dissipation, $\mathcal{L}_{k}\ne0$
and $\overline{\mathcal{L}}_{k}\ne0$, will be demonstrated in Sec.~\ref{sec:Benchmarking-with-Monte}
by comparing the results obtained with the diagrammatic technique
and the data simulated with the Monte Carlo method. The 2D repeater
protocol with the communication time and the temporal filtering protocol
is detailed in Appendix~\ref{Appendix: Full-simulation-of}

\section{Communication time\label{sec:Communication-time}}

In the following section, we incorporate classical communication time
into the diagrammatic technique. For demonstration, we do so for the
1D repeater scheme, which is studied in Sec.~\ref{subsec:1D-repeater}.
The communication time for the 2D repeater protocol is given in Appendix~\ref{Appendix: Full-simulation-of}.

To include the communication time, we take into account that after
each merging attempt, one needs time $t_{m}=t_{\text{swap}}+t_{c}$
in order to continue the generation. Here $t_{\text{swap}}$ is the
time required for a swapping operation, and $t_{c}$ is the time required
for the classical communication between nodes of the segment of current
nesting level $N$ and which scales as $t_{c}\propto2^{N}$. We assume
that the communication time $t_{c}$ is identical for all segments
$\left\{ i\right\} $ at one nesting level. The communication time
does not affect diagrams for the segments generation introduced in
Sec.~\ref{sec:Memories-decay-and} and only changes the diagrams
for the merging.

\paragraph*{Merging of generated segments.}

An extra waiting time $t_{m}$ after a merging operation leads to
an additional memory decoherence $\text{exp}\{\mathcal{L}t_{m}\}$
of the generated state with the Lindblad superoperator $\mathcal{L}$.
The waiting time is incorporated into the diagrammatic technique by
modifying the diagram~\eqref{d3} which represents a successful merging:
\begin{gather}
\diagram{610}\,\rightarrow\,\diagram{10}\label{d6}\\
=\Theta\left(t-t_{m}\right)e^{\mathcal{L}t_{m}}\mathcal{M}\varrho_{\text{prep}}\left(t-t_{m}\right),\nonumber 
\end{gather}
where $\varrho_{\text{prep}}\left(t\right)$ is the density matrix
distribution~\eqref{d2} of the two segments prepared for merging.
Here the delay $t_{m}$ is represented by the Heaviside function $\Theta\left(t-t_{m}\right)$
since there is zero probability to communicate about the merged state
if $t<t_{m}$. The Laplace image of the distribution~\eqref{d6}
is $\text{exp}\{-st_{m}\}\text{exp}\{\mathcal{L}t_{m}\}\mathcal{M}\tilde{\varrho}_{\text{prep}}\left(s\right),$
where $\tilde{\varrho}_{\text{prep}}\left(s\right)$ is defined by
Eq.~$\eqref{eq:rho_prep(s)_1D}$. The Heaviside function in the diagram~\eqref{d6}
results in the extra exponent prefactor. In a similar way, a failed
merging is described by the modified diagram~\eqref{d4} as

\begin{gather*}
\diagram{611}\,\to\,\diagram{305}\\
=\Theta\left(t-t_{m}\right)\text{Tr}\left\{ (\mathcal{I}-\mathcal{M})\varrho_{\text{prep}}\left(t-t_{m}\right)\right\} ,
\end{gather*}
with the Laplace image $\text{exp}\{-st_{m}\}\text{Tr}\left\{ (\mathcal{I}-\mathcal{M})\tilde{\varrho}_{\text{prep}}\left(s\right)\right\} $.
Here we took into account that $\mathcal{L}$ is the trace-preserving
operator.

\paragraph*{Sum over all trajectories.}

The final density matrix distribution is an infinite sum
\begin{gather*}
\varrho(t)=\diagram{306}+\diagram{307}+\diagram{308}+\dots,
\end{gather*}
which in the Laplace domain converges to
\begin{equation}
\tilde{\varrho}\left(s\right)=\frac{e^{\mathcal{L}t_{m}}\mathcal{M}\tilde{\varrho}_{\text{prep}}\left(s\right)}{e^{st_{m}}-\text{Tr}\left\{ (\mathcal{I}-\mathcal{M})\tilde{\varrho}_{\text{prep}}\left(s\right)\right\} }.\label{eq:rho_taret(s)_communication_time}
\end{equation}

Inserting image~$\eqref{eq:rho_taret(s)_communication_time}$ to
Eqs.~$\eqref{eq:-44}$ we obtain the average network state and generation
time

\noindent 
\begin{gather}
\begin{aligned}\rho & =\tilde{\varrho}(s)\Big|_{s=0}=\frac{1}{P}e^{\mathcal{L}t_{m}}\mathcal{M}\tilde{\varrho}_{\text{prep}}\left(0\right)\\
T & =-{\rm Tr}\left\{ \frac{d\tilde{\varrho}(s)}{ds}\right\} \Big|_{s=0}=\frac{1}{P}\left(t_{m}+\frac{3}{2\nu}\right),
\end{aligned}
\label{eq:-7}
\end{gather}

\noindent where $P=\text{Tr}\left\{ \mathcal{M}\tilde{\varrho}_{\text{prep}}\left(0\right)\right\} $
is the average merging probability and $\tilde{\varrho}_{\text{prep}}\left(s\right)$
is defined by Eq.~$\eqref{eq:rho_prep(s)_1D}$. In comparison with
Eqs.~$\eqref{eq:-1}$, where the communication time is not included,
Eqs.~$\eqref{eq:-7}$ take into account the extra decoherence of
the generated state for time $t_{m}$ and include $t_{m}$ to each
cycle of the segment preparation and merging.

\section{Temporal filtering protocol\label{sec:Temporal-filtering-protocol}}

In the section, we develop the diagrammatic technique for the temporal
filtering protocol~\citet{Kuzmin2019}. The temporal filtering protocol
allows one to significantly improve the quality of the entangled states
distributed by the quantum repeaters in the presence of finite memory
times. The idea of the protocol is to discard the state of a prepared
repeater segment as it becomes older than a certain filtering time
$\tau$. After the communication time $t_{c}$, the preparation of
the discarded state is restarted. Although discarding of the prepared
states increases the overall generation time, the protocol suppresses
the effect of the memory decoherence on the final average state generated
by the network.

In fact, the temporal filtering protocol can be considered as the
general case of the repeater protocol~\citet{Briegel1998} and the
so-called “quantum relay” protocol \citet{Jacobs2002}. In the original
quantum repeater protocol, the memory time is required to be much
longer than the time for the state generation, and, therefore, the
temporal filtering is not required, i.e., filtering time $\tau\to\infty$.
In the memory-less quantum relay protocol, all segments states have
to be generated simultaneously, otherwise the prepared states are
discarded, i.e., $\tau\to0$. In contrast to these two extreme cases,
the temporal filtering protocol allows a quantum network to optimally
exploit realistic finite memory coherence times by using a finite
nonzero $\tau$.

Previously, the protocol for re-preparation of the longly decaying
states in 1D repeater was considered in Refs.~\citet{Collins2007,Santra2019,Kuzmin2019}.
In Ref.~\citet{Collins2007}, the segments memories stay unaffected
for the memory time $\tau$, and afterward, the segments states are
re-generated. In Ref.~\citet{Santra2019}, filtration of the segments
states was studied for the 1D repeater with quantum memories subjected
to dephasing. The protocol considered in Ref.~\citet{Santra2019}
also demands that the prepared segments can be used only at discrete
moments of time, therefore, requiring that the generated segments
experience extra unnecessary decoherence while waiting for this time
moments. In the following, we develop the diagrammatic technique which
allows us to consider the temporal filtering protocol for any dissipative
Liouville dynamics of memories, and which can be used in various repeater-network
architectures being free of the mentioned unnecessary waiting time
presented in method~\citet{Santra2019}. Particularly, in our previous
work~\citet{Kuzmin2019}, the developed approach was used for the
2D repeater scheme with the decaying memories.

For illustration, in this section, we describe the diagrammatic technique
incorporating the temporal filtering in the 1D repeater model taking
into account the communication time considered in the previous section.
The temporal filtering does not change the diagrams for the merging
with communication time, it affects only the diagrams for the segments
preparation. The temporal filtering protocol for the 2D repeater scheme
is given in Appendix~\ref{Appendix: Full-simulation-of}.

\paragraph*{Generation of segments within one nesting level.}

In the following we introduce a new diagram representing preparation
of two segments at time $t$ within the time window $[\text{max}\{t-\tau,0\},t].$
There are two possible cases: $t<\tau$ and $t\ge\tau$ which we sum
up as following
\begin{gather}
\diagram{300}=\diagram{14}+\diagram{13}\nonumber \\
=\left[\Theta\left(\tau-t\right)\int_{0}^{t}\bullet\,dt_{0}+\Theta\left(t-\tau\right)\int_{t-\tau}^{t}\bullet\,dt_{0}\right]\nonumber \\
\times p(t)p(t_{0})e^{\mathcal{L}{}_{i}\left(t-t_{0}\right)}\cdot\rho_{\text{in}},\label{eq:-50}
\end{gather}
where we use superoperator notation such that the function after the
square brackets is inserted in place of the bullets. Notice that,
in the second diagram of the sum, with $t\ge\tau$, the integration
over $t_{0}$ is done only within the filtering time $\tau$ preceding
the time moment $t$. The diagram representing two possible orders
for the successful generation of two links with no filtration is
\begin{gather}
\varrho_{{\rm nf}}\left(t\right)=\diagram{300}+\diagram{301}\equiv\diagram{302}.\label{eq:-49}
\end{gather}
After inserting the generation probability distribution~\eqref{eq:-31}
into Eq.~\eqref{eq:-50} one can obtain the Laplace image for Eq.~\eqref{eq:-49}
as

\noindent 
\begin{multline}
\tilde{\varrho}_{{\rm nf}}\left(s\right)=\sum_{ij}\left(1-e^{-\left(s'-\mathcal{L}{}_{i}\right)\tau}\right)\\
\times\frac{\nu^{2}}{\left(s'-\mathcal{L}{}_{i}\right)\left(s'+\nu\right)}\rho_{\text{in}}\Big|_{s'=s+\nu}.\label{eq:-13}
\end{multline}

Now we introduce a diagram representing the probability density to
filter out a segment state of age $\tau$ at time $t$ if the second
segment is not generated
\begin{align}
\diagram{68} & =\Theta\left(t-\tau-t_{c}\right)e^{-\nu(t-t_{c})}p\left(t-\tau-t_{c}\right)\nonumber \\
 & =\Theta\left(t-\tau-t_{c}\right)\nu e^{-\nu\left(2t-2t_{c}-\tau\right)},\label{eq:-52}
\end{align}
with $p(t)$ given in~\eqref{eq:-31}. This diagram is constructed
out of the diagram~\eqref{d1} describing a successful generation
of a segment at time $t-\tau-t_{c}$ and a vertical line of length
$t-t_{c}$ representing the unsuccessful attempts to prepare the second
segment during time $t-t_{c}$. In the diagram~\eqref{eq:-52}, we
take into account that, as the segment is discarded, the communication
time $t_{c}$ is required to inform the nearby nodes that the link
has to be regenerated as it grew too ``old''. Here, we neglect a small
probability $1-\text{exp}(-\nu t_{c})\approx\nu t_{c}$, that the
unprepared link can be generated within the communication time $t_{c}\ll1/\nu$,
and, therefore, in the diagram, its generation is also stopped for
the communication time $t_{c}$. This assumption is required in order
to sum up the diagram series, and it only slightly overestimates the
effect of the communication time. This will be verified below by benchmarking
against the MC method.

The probability distribution for the two possible events of the segments
filtration is represented by the following diagram

\begin{gather}
p_{{\rm f}}\left(t\right)=\diagram{67}+\diagram{68}\equiv\diagram{69}\label{eq:-53}
\end{gather}
with the Laplace image obtained from Eq.~\eqref{eq:-52} as
\begin{equation}
\tilde{p}_{{\rm f}}\left(s\right)=2e^{-s\left(\tau+t_{c}\right)}\frac{\nu e^{-\nu\tau}}{s+2\nu}.\label{eq:-4}
\end{equation}

Using diagrams~\eqref{eq:-49} and \eqref{eq:-53} we obtain the
density matrix distribution for the two segments prepared to be merged:

\begin{gather}
\varrho_{\text{prep}}\left(t\right)=\diagram{302}+\diagram{303}+\diagram{304}+\dots,\label{eq:-54}
\end{gather}
which, in the Laplace domain, reads

\noindent 
\begin{align}
\tilde{\varrho}_{\text{prep}}\left(s\right) & =\sum_{n=0}^{\infty}\tilde{\varrho}_{{\rm nf}}\left(s\right)\big[\tilde{p}_{{\rm f}}\left(s\right)\big]^{n}=\frac{\tilde{\varrho}_{{\rm nf}}\left(s\right)}{1-\tilde{p}_{{\rm f}}\left(s\right)}.\label{eq:-55}
\end{align}

\paragraph*{Merging and sum over all trajectories.}

\noindent The calculation of the density matrix distribution for the
merged segment is identical to what was presented in Sec.~\ref{sec:Communication-time}.
Therefore, one can obtain the target Laplace image $\tilde{\varrho}\left(s\right)$
by inserting $\tilde{\varrho}_{\text{prep}}\left(s\right)$ from the
Eq.~\eqref{eq:-55} into the Eq.~$\eqref{eq:rho_taret(s)_communication_time}$.
The average generated state and the generation time are obtained using
Eqs.~$\eqref{eq:-44}$ as usual:

\noindent 
\begin{gather}
\begin{aligned}\rho & =e^{\mathcal{L}t_{m}}\frac{\mathcal{M}\tilde{\varrho}_{\text{prep}}\left(0\right)}{\text{Tr}\left\{ \mathcal{M}\tilde{\varrho}_{\text{prep}}\left(0\right)\right\} }=\frac{1}{2PP_{{\rm nf}}}e^{\mathcal{L}t_{m}}\times\\
 & \times\left[\frac{1-e^{-\left(1-\mathcal{L}{}_{1}/\nu\right)\nu\tau}}{1-\mathcal{L}{}_{1}/\nu}+\frac{1-e^{-\left(1-\mathcal{L}{}_{2}/\nu\right)\nu\tau}}{1-\mathcal{L}{}_{2}/\nu}\right]\rho_{\text{in}},\\
T & =\frac{1}{P}\left(t_{m}+\frac{3}{2\nu}+\frac{t_{c}+\frac{1}{2\nu}}{e^{\nu\tau}-1}\right).
\end{aligned}
\label{eq:-42}
\end{gather}

\noindent 

\noindent Here we have introduced the probability of not filtering
the segment state $P_{{\rm nf}}=1-\tilde{p}_{{\rm f}}\left(0\right)=1-e^{-\nu\tau}$,
the merging probability $P=\text{Tr}\left\{ \mathcal{M}\tilde{\varrho}_{\text{prep}}\left(0\right)\right\} $,
and the time $t_{m}=t_{c}+t_{\text{swap}}$ required for the merging
including classical communication delay.

\begin{figure}
\includegraphics{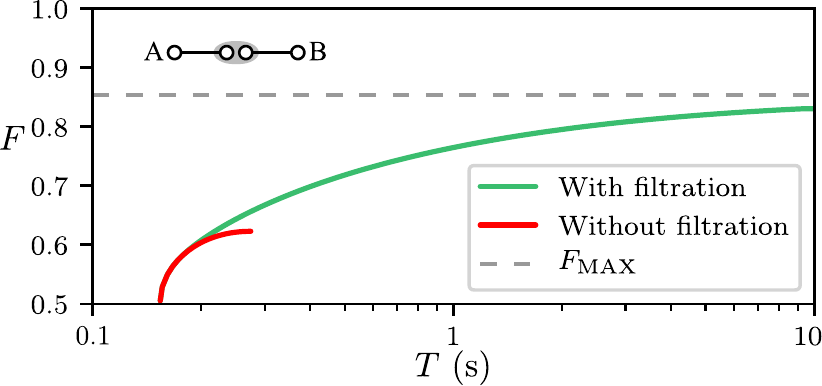} \caption{Maximum fidelity $F=\left\langle \psi\right\vert \rho\left\vert \psi\right\rangle $
versus the generation time $T$ achievable in the 1D DLCZ repeater
of the nesting level I distributing the Bell state $\left\vert \psi\right\rangle =\left(\left|01\right\rangle +\left|10\right\rangle \right)/\sqrt{2}$
among two parties A and B, as shown in the inset. Network parameters:
length $L$ = 300 km, memory decay time $T_{\text{coh}}=100$ ms,
and other experimental parameters specified in \citet{Parameters}.
Green (red) line represents optimized results obtained with (without)
temporal filtering. The gray dashed line indicates fidelity $F_{\text{MAX}}$
achievable with the network in the limit $T_{\text{coh}}=\infty$.}
\label{fig:-3}
\end{figure}

\paragraph*{Example of a network with filtration.}

The striking feature of the temporal filtering protocol is that it
can enable applications requiring entangled states of higher quality
than the standard repeater protocol can generate given a finite memory
coherence time. As presented in the Fig.~$\ref{fig:-3}$, the temporal
filtering protocol allows one to trade the generation rate~$1/T$
for increase in the fidelity~$F$ of the resulting states by decreasing
the filtering time~$\tau$. The red curve is obtained by only optimizing
the elementary segments generation (i.e. squeezing parameter~$\epsilon$
in the DLCZ protocol, see the next section) without using the temporal
filtering ($\tau=\infty$). For the green curve, the filtering time~$\tau$
was numerically optimized. The plot illustrates that the filtering
allows the network to achieve much higher fidelities. Particularly,
in the given example with memory decay time~$T_{\text{coh}}=100$
ms and network length $L=300$ km, using filtration one can almost
reach the maximum fidelity $\approx0.85$ achievable in the limit
of infinite coherence time $T_{\text{coh}}\to\infty$, while increasing
the generation time to acceptable $T\approx10$ s. On the other hand,
the highest achievable fidelity without filtering is $\approx0.62$.

As a result, with the temporal filtering protocol, the same quantum
repeater network can be exploited in two regimes: high rate regime,
which could be used for the secret key distribution, and high fidelity
regime, which could be used for the distributed quantum computation.
Notice that, in a real-world network, the switch between two regimes
can be implemented in real-time by varying the filtering parameter
$\tau$.

\section{Benchmarking against Monte Carlo simulation\label{sec:Benchmarking-with-Monte}}

\begin{figure}
\includegraphics{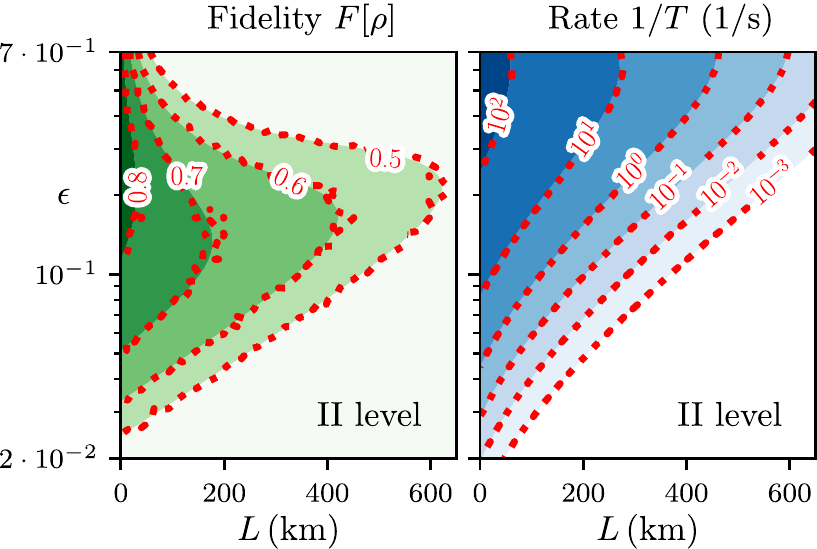}

\caption{Fidelity $F=\left\langle \psi\right\vert \rho\left\vert \psi\right\rangle $
(left panel) and generation rate $1/T$ (right panel) of the states
generated with the 1D DLCZ repeater of the nesting level II {[}see
Fig. $\ref{fig:1}$(a){]} versus network length $L$ and squeezing
parameter $\epsilon$ in generation of the elementary segments with
the DLCZ protocol (see Sec.~\ref{sec:Methods-and-results} and Appendix~\ref{Appendix: Calculation-of-the}).
Network parameters: memory decay time $T_{\text{coh}}=150$ ms, filtering
time $\tau=15$ ms, and other experimental parameters specified in
\citet{Parameters}. Color contours represent data obtained by Eqs.~$\eqref{eq:-42}$,
derived with the diagrammatic technique. Red dotted contours give
data simulated with the Monte-Carlo (MC) method, such that each data
point is obtained averaging over $4151$ trajectories. Red numbers
present data values for the contours.}
\label{fig:Benchmark_verification}
\end{figure}

To assess the accuracy of the diagrammatic technique we compare results
obtained for the 1D DLCZ repeater of the nesting level II {[}see Fig.
$\ref{fig:1}$(a){]} using semi-analytical method given by Eqs.~$\eqref{eq:-42}$
with the results of the MC simulation, serving as a benchmark. Details
on the MC simulation are presented in~\citep{Kuzmin2019}.

In the original one-dimensional DLCZ repeater, the elementary segments
are generated as follows. Weak coherent pulses drive distant atomic
ensembles (memories) resulting in entangled two-mode squeezed states
of the ensembles and the out-coming photon modes $\sum_{n=0}^{\infty}\epsilon^{n}\left\vert n\right\rangle _{m}^{(i)}\left\vert n\right\rangle _{p}^{(i)}$
with squeezing parameter $\epsilon\ll1$; here the indexes $m$ and
$p$ correspond to the memory and the photon modes and $i\in\{1,2\}$
refers to the memory index. The two out-coming photon modes interfere
with each other on a beam-splitter placed in-between of the ensembles
and are measured afterward. Upon a probabilistic detection of a single
photon, the state of the memories is projected onto an entangled state
spanning the elementary segment length $L_{0}$. A probabilistic swapping
operation of two prepared segments is implemented by mapping the states
of the neighboring memories to photon pulses, which after interfering
on a beam-splitter are measured. The swapping is successful upon a
single-photon detection. The details on the simulation of the elementary
segments in the DLCZ scheme are given in Appendix~\ref{Appendix: Calculation-of-the}.

We consider experimental parameters realistic for current room-temperature
atomic ensembles~\citet{Borregaard2015,Katz2018}: memory decay time
$T_{\text{coh}}=150$ ms and the length of the signal pulses $10^{-4}s$,
which we use as $t_{\text{swap}}$ — the time for the merging operation.
The filtering time $\tau=15$ ms is chosen such that the temporal
filtering protocol significantly limits the degradation of the network
memories. Other experimental parameters are specified in \citet{Parameters}.

The simulation results are presented in the Fig~$\ref{fig:Benchmark_verification}$.
Color contours present the fidelity of the prepared states with respect
to the target Bell state (left panel) and the average generation rate
(right panel) as functions of the linear network size $L$ and the
squeezing parameter $\epsilon$. The results of the MC simulation
are given by red dotted contours and perfectly match the data obtained
with the semi-analytical equations~\eqref{eq:-42}. Thereby, the
Fig~$\ref{fig:Benchmark_verification}$ verifies the accuracy of
the assumptions made to develop the diagrammatic method. On the other
hand, the Fig.~\ref{fig:-2} demonstrates a dramatic advantage of
the proposed semi-analytical approach over the MC simulation in the
calculation performance.

\section{Application for Network optimization\label{sec:Network-optimization}}

Due to the high accuracy and the small calculation run-time, the developed
technique is a powerful tool for optimization of real-world large-scale
quantum networks. For example, in the Fig.~\ref{fig:Benchmark_verification}(left
panel) one can see that there is an optimal squeezing parameter $\epsilon$
which maximizes the entanglement distribution distance for the DLCZ-type
repeater. Note that the Fig.~\ref{fig:Benchmark_verification} represents
results of the simulation for a comparably small 1D network of the
nesting level II because it has to be accessible for the MC simulation
method.

\begin{figure}
\includegraphics{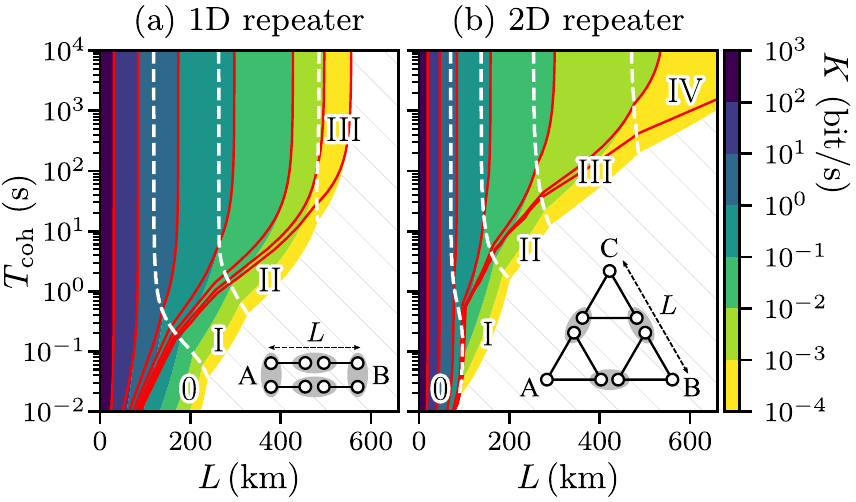}\caption{The optimized rate $K$ of the secret key distribution (a) between
two parties with the 1D DLCZ repeater and (b) among three parties
with the 2D repeater, obtained with the semi-analytical technique.
The network schemes are sketched in the insets. Both panels display
maximum rate $K$ versus memory coherence time $T_{\text{coh}}$ and
covered distance $L$ for the experimental parameters specified in~\citep{Parameters}.
The parameters include communication time and time for the signal
detection $10^{-4}s$ in the merging operations, realistic for the
atomic ensemble~\citet{Borregaard2015}. Filled color contours show
results for the repeaters with the temporal filtering protocol, where
white dashed contours separate areas with different optimal nesting
levels numerated by the Roman numbers. Red contours indicate results
of optimization for the repeaters without the temporal filtering protocol.}
\label{fig:-2-1}
\end{figure}

Below we give examples of simulations of much larger networks, including
a 2D network of the nesting level IV, which are made possible by the
semi-analytic technique. In particular, we optimize quantum networks
to maximize the rate of the secret key~\citet{Epping2017} distribution
among two parties with the 1D DLCZ repeater network and among three
parties with the 2D repeater network. The maximization of the secret
key rate, $K$, is of a particular interest because $K$ is a function
of both, the quality and the generation rate of the prepared entangled
states {[}see Appendix~\ref{Appendix: Secret-key-rate}{]}, which,
in the probabilistic repeater protocols, are traded one for another.
For the demonstration, we consider the protocols with and without
the temporal filtering. In the latter case, we optimize over the network
nesting level and the squeezing parameter $\epsilon$, and in the
former case, the filtering time $\tau$ is added as the third parameter
to optimize. The details on the simulation of the elementary segments
in the 2D repeater scheme are given in Appendix~\ref{Appendix: Calculation-of-the}.

Figure~$\ref{fig:-2-1}$ shows the maximum rate $K$ of the secret
key distribution as a function of the distribution length $L$ and
the memory decay time $T_{\text{coh}}$. The left panel corresponds
to the 1D network and the right panel shows results for the 2D network.
The plots highlight the benefits of using quantum networks with higher
nesting levels to cover longer distances in the presence of the sufficiently
long memory coherence time. One can also see that the temporal filtering
protocol~(filled color contours) provides higher key rates than the
no filtering strategy~(red contours) in the domain of small memory
coherence times. Remarkably, this is achieved without usage of extra
physical resources (e.g. multiplexing) but only with optimization
of the filtering time $\tau$. The given example of the networks optimizations
illustrates the potential of the developed method for efficient designing
and optimization of the real-world large-scale quantum internet.

\section{Conclusion\label{sec:Outlook}}

We have presented in details and developed further the diagrammatic
technique used in~\citet{Kuzmin2019} to study 2D repeater networks.
We showed examples of using the technique to evaluate the average
states and the average generation times of the 1D and 2D repeater
networks taking into account the communication time and incorporating
the temporal filtering protocol. Compared to other analytical methods,
our technique accounts for the continuous dissipative Liouville dynamics
of the network quantum memories. The results obtained with the semi-analytic
technique match the exact MC simulations while the required computational
resources scales only linearly with the network size. We stress that
the presented technique is not limited to the considered repeater
models but can be applied to evaluate networks with more complex configurations
realized on various experimental platforms. The diagrammatic technique
is a convenient tool for investigation of new repeater protocols as
it allows for precise comparison of the protocols subjected to relevant
realistic imperfections. The method can be useful for designing and
optimization of future real-world quantum networks.
\begin{acknowledgments}
We thank C. A. Muschik and W. Dür for fruitful discussions on the
quantum repeater networks. Research was sponsored by the Army Research
Laboratory and was accomplished under Cooperative Agreement Number
W911NF-15-2-0060. The views and conclusions contained in this document
are those of the authors and should not be interpreted as representing
the official policies, either expressed or implied, of the Army Research
Laboratory or the U.S. Government. The U.S. Government is authorized
to reproduce and distribute reprints for Government purposes notwithstanding
any copyright notation herein.
\end{acknowledgments}

\appendix

\section{Calculation of the elementary segments and the merging superoperator\label{Appendix: Calculation-of-the}}

\begin{figure}
\includegraphics{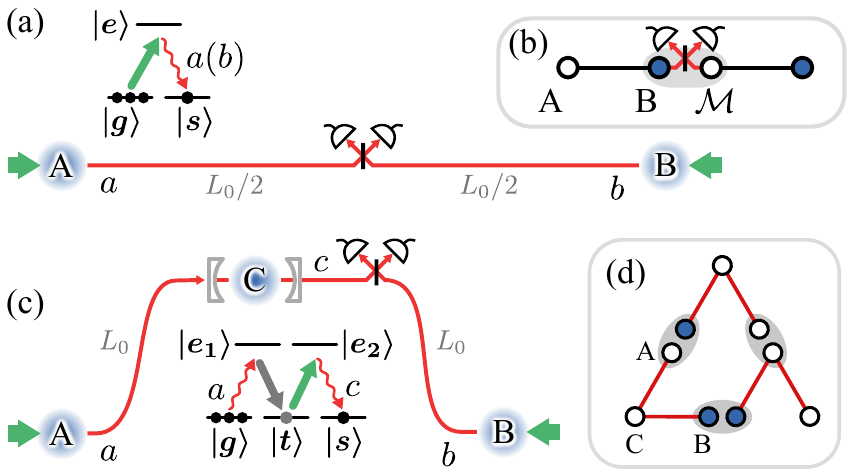} \caption{Implementation schemes of the elementary segments of (a) the 1D~\citet{Duan2001}
and (c) the 2D~\citet{Kuzmin2019} repeaters. Blue gradient filled
circles represent atomic ensembles with level structures depicted
nearby. Bold arrows depict weak classical fields, red thin arrows
represent photonic modes, half-circles represent single-photon detectors.
In (c), node C is placed in a high-finesse cavity. (b) Quantum swapping
operation for two prepared segments. The states of the nodes, connected
by links, of the same (different) colors are correlated (anti-correlated).
(d) A possible relevant position of the elementary segments of the
2D repeater for merging to nesting level I with three swapping operations
depicted in (b). Red links correspond to the photonic modes as in
(c).}
\label{fig:-5}
\end{figure}

In this section, we describe simulation of the elementary segments
in the 1D and 2D quantum repeater protocols. For each protocol, we
calculate the state of the elementary segment, $\rho_{e}$, success
probability $q$ to generate the segment for one attempt, and time
$\Delta t$ required for one generation attempt. We also calculate
the merging superoperator $\mathcal{M}$, taking into account all
relevant experimental imperfections.

\subsection*{1D DLCZ protocol}

The scheme of the 1D DLCZ protocol~\citet{Duan2001} is shown in
Fig.~\ref{fig:-5}(a). Two nodes of the segment, A and B, are distributed
for a distance $L_{0}$. Each node consists of an atomic ensemble
with a $\Lambda$-type level structure. Information in the ensembles
is encoded in the absence or presence of a collective spin excitation,
i.e., in the the logical states $|0\rangle=|g_{1}g_{2}...g_{N}\rangle$
and $|1\rangle=\frac{1}{\sqrt{N}}\sum_{i=1}^{N}|g_{1}g_{2}...s_{i}...g_{N}\rangle$,
where $N$ is the number of emitters per ensemble, and $\left\vert g_{i}\right\rangle $,
$\left\vert s_{i}\right\rangle $ denote the levels of the emitters.
The target state of the elementary segment is $(\left\vert 1_{A}0_{B}\right\rangle +\left\vert 0_{A}1_{B}\right\rangle )/\sqrt{2}$,
where subscripts refer to the corresponding ensembles. The generation
of the elementary segment is probabilistic and is performed in discrete
steps. At the beginning of each step, all ensembles are prepared in
the states $\left\vert 0\right\rangle $. Then, weak laser pulses
drive the ensembles at nodes A and B, coupling them to the out-going
photonic modes, $a$ and $b$ correspondingly, via off-resonant Raman
transition. The resulting entangled states of the ensembles and the
photonic modes are two two-mode squeezed vacuum states 
\begin{multline}
\left\vert \Psi_{Aa\left(Bb\right)}\left(\epsilon_{a(b)}\right)\right\rangle =\text{sech}\,r_{a(b)}\\
\times\sum_{n=0}^{\infty}\left(\text{tanh}\,r_{a(b)}\right)^{n}\left\vert n_{A\left(B\right)}\right\rangle \left\vert n_{a\left(b\right)}\right\rangle \\
\propto\sum_{n=0}^{\infty}\epsilon_{a(b)}^{n}\left\vert n_{A\left(B\right)}\right\rangle \left\vert n_{a\left(b\right)}\right\rangle .\label{eq:-13-1}
\end{multline}
Here $r_{a(b)}$ is the squeezing parameter, which is controlled by
the strength of the driving laser. For simplicity, we use a notation
$\epsilon_{a(b)}\equiv\text{tanh}\,r_{a(b)}\in[0,1)$.

Each of the outgoing photonic modes $i=\{a,b\}$ propagates through
the length $L_{0}/2$ of lossy fiber. Fiber losses are represented
by a loss superoperator $\mathcal{S}_{\text{loss}}^{i}\left(L_{0}/\left[2L_{\text{att}}\right]\right)$,
where
\begin{equation}
\mathcal{S}_{\text{loss}}^{i}\left(g\right)\bullet=\text{exp}\left\{ g\,\mathcal{D}\left[a_{i}\right]\right\} \bullet,\label{eq:-28}
\end{equation}
with $a_{i}$ the annihilation operator in the mode $i$, $\mathcal{D}[a]\bullet=a\bullet a^{\dagger}-\frac{1}{2}\left(a^{\dagger}a\bullet+\bullet a^{\dagger}a\right)$
the Lindblad superoperator, and $L_{\text{att}}$ the attenuation
length. Afterward, both photonic modes enter the station for the probabilistic
swapping operation placed in-between of the two nodes, and which consists
of a perfect balanced beamsplitter and two imperfect single-photon
detectors. The photonic modes interfere in the beamsplitter and the
outgoing modes are measured. Probabilistic detection of a single photon
heralds generation of an entangled state, otherwise the generation
attempt is repeated. Detectors imperfections include dark counts with
rate $d$ and a photon loss with probability $f$. The latter is represented
by loss superoperator~$\eqref{eq:-28}$ and the former by a dark
counts superoperator 
\[
\mathcal{S}_{\text{dc}}^{i}\left(d\right)\bullet=\text{exp}\left\{ d\,\mathcal{D}[a_{i}]+d\,\mathcal{D}[a_{i}^{\dagger}]\right\} \bullet.
\]
A superoperator for the full swapping operation acting to two photonic
modes $i$ and $j$ is 
\begin{equation}
\mathcal{M}_{s}^{ij}\bullet=2\left\langle 1_{i}0_{j}\right\vert \{(\mathcal{S}_{\text{det}}^{i}\otimes\mathcal{S}_{{\rm det}}^{j})\mathcal{S}_{\text{BS}}^{ij}\bullet\}\left\vert 1_{i}0_{j}\right\rangle ,\label{eq:-29}
\end{equation}
where $\mathcal{S}_{\text{det}}^{i}=\mathcal{S}_{\text{dc}}^{i}\left(d\right)\mathcal{S}_{\text{loss}}^{i}\left(-\text{ln}[1-f]\right)$.
Projection $\left\langle 1_{i}0_{j}\right\vert \bullet\left\vert 1_{i}0_{j}\right\rangle $
describes detection of a single photon in the photonic mode $i$.
The prefactor $2$ accounts for the possibility to detect a photon
in the second photonic mode $j$ with the corresponding projector
$\left\langle 0_{i}1_{j}\right\vert \bullet\left\vert 0_{i}1_{j}\right\rangle $.
Photon detection in different detectors give opposite phase of the
resulting states which can be reduced to each other by the phase flip
operation. In Eq.~\eqref{eq:-29}, the superoperator for the balanced
beamsplitter reads
\begin{align*}
\mathcal{S}_{\text{BS}}^{ij} & \bullet=U_{{\rm BS}}^{ij}\bullet\left(U_{{\rm BS}}^{ij}\right)^{\dagger},\\
U_{{\rm BS}}^{ij} & =\text{exp}\left[\frac{\pi}{4}\left(a_{i}^{\dagger}a_{j}-a_{i}a_{j}^{\dagger}\right)\right].
\end{align*}
Swapping operation $\eqref{eq:-29}$ is not trace-preserving, and
thus is probabilistic with the success probability defined by the
trace of the resulting state as $\text{Tr}\{\mathcal{M}_{s}^{ij}\bullet\}$.

The success/failure of the swapping operator is communicated from
the central station to the nodes of the segment. Therefore, time for
one generation attempt $\Delta t=v_{\text{c}}L_{0}+t_{s}$ comprises
time $v_{\text{c}}L_{0}/2$ required for the signals to propagate
from the nodes to the swapping station, time $t_{s}$ for the the
signal detection, and time $v_{\text{c}}L_{0}/2$ for the back communication.
Here, $v_{\text{c}}$ is the speed of light in the fiber and $t_{s}$
is the photon pulse duration. During time $\Delta t$, memories of
the nodes A and B experience decay accounted by applying the loss
superoperator~\eqref{eq:-28} $\mathcal{S}_{\text{loss}}^{i}\left(\Delta t/T_{\text{coh}}\right)$
to both memory modes $i=\{A,B\}$, where $T_{\text{coh}}$ is the
memory coherence time.

In the following, we will use the matrix-vector representations of
states~$\eqref{eq:-13-1}$ $|\rho_{Jj}\left(\epsilon_{j}\right)\rangle\rangle\equiv\left\vert \Psi_{Jj}\left(\epsilon_{j}\right)\right\rangle \left\langle \Psi_{Jj}\left(\epsilon_{j}\right)\right|$
with $Jj=\{Aa,Bb\}$. In the matrix-vector representations, generated
state $\rho_{e}$ of the 1D repeater elementary segment reads
\begin{multline}
|\rho_{e}\left(\epsilon\right)\rangle\rangle=\mathcal{M}_{s}^{ab}\\
\times\underset{Jj=\{Aa,Bb\}}{\prod\otimes}\mathcal{S}_{\text{loss}}^{J}\left(\frac{\Delta t}{T_{\text{coh}}}\right)\mathcal{S}_{\text{loss}}^{j}\left(\frac{L_{0}}{2L_{\text{att}}}\right)|\rho_{Jj}\left(\epsilon\right)\rangle\rangle,\label{eq:-39}
\end{multline}

\noindent where we use the optimal relation of the parameters $\epsilon_{a}=\epsilon_{b}\equiv\epsilon$.
The success probability of one generation attempt is given by 
\[
q=\text{Tr}\left\{ \rho_{e}\left(\epsilon\right)\right\} .
\]
Free parameter $\epsilon$ is numerically optimized for each set of
the network parameters to obtain the results illustrated in Figs.~\ref{fig:-3}
and \ref{fig:-2-1} of the main text.

The merging operation of two prepared segments is presented in Fig.~\ref{fig:-5}(b):
states of two memories are read-out and directed to the swapping station,
which is identical to the swapping station used in the preparation
of the elementary segment. Ideally, the detection of a single photon
projects the joint system into the entangled state. Taking into account
inefficiency $v$ of the read-out operation and the imperfect detectors,
one can represent the merging operator acting to two modes $i$ and
$j$ via superoperators~\eqref{eq:-28} and \eqref{eq:-29} as
\begin{equation}
\mathcal{M}\bullet=\mathcal{M}_{s}^{ij}\mathcal{S}_{\text{loss}}^{i}\left(-\text{ln}[1-v]\right)\mathcal{S}_{\text{loss}}^{j}\left(-\text{ln}[1-v]\right)\bullet.\label{eq:-35}
\end{equation}
The derived merging operator $\mathcal{M}$ is used in the numerical
simulations presented in the paper.

For $\epsilon\ll1$ only several excited states of the quantum memories
and the photon modes are considerably populated. Therefore, we truncate
dimensions of the bosonic modes to the Fock state $\left\vert 2\right\rangle $,
thus, taking into account the first order of the multi-excitation
error.

\subsection*{2D repeater protocol}

In the subsection, we explain the simulation of the elementary segments
of the 2D repeater protocol~\citet{Kuzmin2019}. The scheme of the
elementary segment is shown in Fig.~\ref{fig:-5}(c). The segment
consists of three nodes A, B, and C, which are distributed in space
such that the scheme has a shape of a triangle (equilateral for simplicity)
with the length $L_{0}$ of the sides. Similarly to the nodes of the
1D scheme, nodes A and B contain atomic ensembles with a $\Lambda$-type
level structure. Node C employs a cold atomic ensemble with a double-$\Lambda$
configuration placed in a cavity with good cooperativity.

The ideal target state of the elementary segment is $\left\vert \Psi_{\text{GHZ}}\right\rangle =(\left\vert 1_{A}1_{C}0_{B}\right\rangle +\left\vert 0_{A}0_{C}1_{B}\right\rangle )/\sqrt{2}$.
The generation of the elementary segment is probabilistic and is performed
in discrete steps. At the beginning of each step, atoms in the ensembles
are pumped into the states $\left|g\right\rangle $. Then, nodes A
and B are driven with weak laser pulses to produce two-mode squeezed
vacuum states $\eqref{eq:-13-1}$, similar to the 1D repeater described
above. The photonic modes propagate through the distance $L_{0}$
via lossy fiber. This is modeled by applying the loss superoperators~\eqref{eq:-28}
$\mathcal{S}_{\text{loss}}^{i}\left(L_{0}/L_{\text{att}}\right)$
to each photonic mode.

Mode $a$ is directed to node C designed to implement a nonlinear
gate $\left\vert 1_{a}0_{C}0_{c}\right\rangle \rightarrow\left\vert 0_{a}1_{C}1_{c}\right\rangle $,
$\left\vert 0_{a}0_{C}0_{c}\right\rangle \rightarrow\left\vert 0_{a}0_{C}0_{c}\right\rangle $.
Such an evolution is given by a unitary $U=\text{exp}[\frac{\pi}{2}(aC^{\dagger}c^{\dagger}-\text{h.c.})]$,
which describes swapping of the incoming photon in mode $a$ into
one outgoing photon in mode $c$ and one excitation in the collective
spin mode $C$ of the ensemble. The inefficiency of the gate is modeled
as a loss of excitation in the incoming mode $a$ and is represented
by the loss superoperator $S_{\text{loss}}^{a}(-\text{ln}\,\eta)$
given by Eq.~\eqref{eq:-28}, where $\eta$ is the gate efficiency.
As a result, the evolution in node~B is given by the superoperator
\[
\mathcal{S}_{C}\bullet=\text{Tr}_{a}\left\{ U\left[S_{\text{loss}}^{a}(-\text{ln}\,\eta)\,\bullet\right]U^{\dagger}\right\} ,
\]
where the incoming mode $a$ is traced out since it is not measured
after the interaction.

Afterward, photonic modes $b$ and $c$ are directed into the swapping
station placed right after node C. The swapping station is similar
to the one used in the 1D scheme, and is described by the superoperator~\eqref{eq:-29}
$\mathcal{M}_{s}^{ac}$.

The time for one generation attempt $\Delta t=2v_{\text{c}}L_{0}+t_{s}$
consists of the time required for the propagation of the signals to
the swapping station, the time $t_{s}$ for the signal detection in
the swapping operation, and the classical communication time. Thus,
the memories in nodes A and B experience decay for time $\Delta t$
and the memory in node C decays for time $T_{C}=v_{\text{c}}L_{0}+t_{s}/T_{\text{coh}}$.
The resulted state of the elementary segment reads
\begin{multline*}
|\rho_{e}\left(\epsilon_{a},\epsilon_{b}\right)\rangle\rangle=\mathcal{M}_{s}^{ac}\mathcal{S}_{\text{loss}}^{C}\left(\frac{T_{C}}{T_{\text{coh}}}\right)\mathcal{S}_{C}\\
\times\underset{Jj=\{Aa,Bb\}}{\prod}\mathcal{S}_{\text{loss}}^{J}\left(\frac{\Delta t}{T_{\text{coh}}}\right)\mathcal{S}_{\text{loss}}^{j}\left(\frac{L_{0}}{2L_{\text{att}}}\right)|\rho_{Jj}\left(\epsilon_{j}\right)\rangle\rangle\\
\otimes|0_{C}0_{c}\rangle\rangle.
\end{multline*}
The success probabilities for one generation attempt is 
\[
q=\text{Tr}\left\{ \rho\left(\epsilon_{a},\epsilon_{b}\right)\right\} .
\]

\noindent Free parameters $\epsilon_{a(b)}$ are numerically optimized
for each set of the network parameters to obtain the results illustrated
in Fig.~\ref{fig:-2-1} of the main text.

Fig.~\ref{fig:-5}(d) presents an example of a possible relative
orientation of three elementary segments in the 2D network. This configuration
is used for numerical analysis of the 2D repeater shown in Fig.~\ref{fig:-2}.
For simulations of both, 1D and 2D repeaters, we truncate the Hilbert
space of the bosonic modes up to the Fock state $\left\vert 2\right\rangle $.

\section{Full simulation of the 2D repeater\label{Appendix: Full-simulation-of}}

In the following section, we use the diagrammatic technique to describe
the two-dimensional (2D) repeater~\citet{Kuzmin2019} presented in
Fig.~$\ref{fig:1}$(b). Here we develop further the results of Subsec.~\ref{subsec:2D-repeater}
of the main text by incorporating the communication time and the temporal
filtering protocol presented, correspondingly, in Sections~\ref{sec:Communication-time}
and \ref{sec:Temporal-filtering-protocol} of the main text for the
1D repeater.

Unlike the temporal filtering protocol for the 1D repeater, the 2D
repeater version exploits two filtering times: $\tau_{1}$ – time,
after which the state of the first prepared segment $i$ is discarded;
and $\tau_{2}$ – time, after which we discard the state resulted
from the merging of the segments $i$ and $j$. First, we can reuse
diagram~\eqref{eq:-49} of the main text to represent the successful
generation of two segments, $i$ and $j$, with no time filtering
as 
\[
\varrho_{{\rm nf},k}^{(1)}\left(t\right)=\diagram{500},
\]
such that the third segment, $k$, is unprepared. In the Laplace domain
the extra bar results in a frequency shift of the Laplace image~\eqref{eq:-13}
by $\nu$

\begin{align*}
\tilde{\varrho}_{{\rm nf},k}^{(1)}\left(s\right) & =\sum_{ij\ne k}\left(1-e^{-\left(s'-\mathcal{L}{}_{i}\right)\tau}\right)\\
 & \times\frac{\nu^{2}}{\left(s'-\mathcal{L}{}_{i}\right)\left(s'+\nu\right)}\rho_{\text{in}}\Big|_{s'=s+2\nu}.
\end{align*}

Similarly, the probability distribution for all possible events of
the segments filtration can be obtained by adding a bar to the diagram~\eqref{eq:-52}
and summing over filtration of each of three segments
\[
\sum_{i}\diagram{41}\equiv\diagram{42}
\]
 The Laplace image of this diagram is
\[
\tilde{p}_{{\rm f}}^{(1)}\left(s\right)=3e^{-s\left(\tau_{1}+t_{c}\right)}\frac{\nu e^{-2\nu\tau}}{s+3\nu}.
\]

All possible trajectories for preparation segments $i$ and $j$ are
represented as
\[
\varrho_{\text{prep},k}^{(1)}\left(t\right)=\diagram{500}+\diagram{501}+\diagram{502}+...,
\]
with the Laplace image
\[
\tilde{\varrho}_{\text{prep},k}^{(1)}\left(s\right)\equiv\frac{\tilde{\varrho}_{{\rm nf},k}^{(1)}\left(s\right)}{1-\tilde{p}_{{\rm f}}^{(1)}\left(s\right)}.
\]

We denote the successful attempt of merging of prepared segments $i$
and $j$ with unprepared segment $k$ as 
\begin{equation}
\diagram{47}=\Theta\left(t-t_{m}\right)e^{-\mathcal{L}_{ij}t_{m}}\mathcal{M}{}_{k}\varrho_{\text{prep},k}^{(1)}\left(t\right),\label{eq:-3}
\end{equation}
where the superoperator $\mathcal{L}_{ij}$ describes degradation
of the resulted state for the time $t_{m}$ required by the merging.
All possible variants of the unsuccessful merging attempts are denoted
as
\begin{equation}
\diagram{48}=\Theta\left(t-t_{m}\right)\text{Tr}\left\{ \sum_{k}\left(\mathcal{\mathcal{I}}-\mathcal{M}{}_{k}\right)\varrho_{\text{prep},k}^{(1)}\left(t\right)\right\} ,\label{eq:-5}
\end{equation}
where $k\in\{1,2,3\}$. Then, the density matrix distribution for
generation of two segments is the following sum of trajectories

\[
\varrho_{k}^{(2)}(t)=\diagram{49}+\diagram{50}+\diagram{51}+...\equiv\diagram{620},
\]
with the Laplace image obtained using Eqs.~\eqref{eq:-3},~\eqref{eq:-5}

\[
\tilde{\varrho}_{k}^{(2)}(s)=\frac{e^{-\mathcal{L}_{ij}t_{m}}\mathcal{M}{}_{k}\tilde{\varrho}_{\text{prep},k}^{(1)}\left(s\right)}{e^{st_{m}}-\text{Tr}\left\{ \sum_{k}\left(\mathcal{\mathcal{I}}-\mathcal{M}_{k}\right)\tilde{\varrho}_{\text{prep},k}^{(1)}\left(s\right)\right\} }.
\]

Generation of the third segment $k$ within the time period $\tau_{2}$
after the merging of segments $i$ and $j$ is represented as

\begin{align*}
\varrho_{{\rm nf},k}^{(2)}\left(t\right) & =\diagram{621}+\diagram{622}\\
 & =\left[\Theta\left(\tau_{2}-t\right)\int_{0}^{t}\bullet dt_{0}+\Theta\left(t-\tau_{2}\right)\bullet\int_{t-\tau_{2}}^{t}dt_{0}\right]\\
 & \times p_{k}\left(t-t_{0}\right)e^{\mathcal{L}{}_{ij}\left(t-t_{0}\right)}\varrho_{k}^{(2)}(t_{0}),
\end{align*}
with the Laplace image

\noindent 
\[
\tilde{\varrho}_{{\rm nf},k}^{(2)}\left(s\right)=\left(1-e^{-\left(s+\nu-\mathcal{L}{}_{ij}\right)\tau_{2}}\right)\frac{\nu}{s+\nu-\mathcal{L}{}_{ij}}\tilde{\varrho}_{k}^{(2)}(s).
\]

All variants for the filtration of the merged state are represented
as

\begin{gather*}
p_{{\rm f}}^{(2)}\left(t\right)=\sum_{k}\diagram{623}\\
=\Theta\left(t-\tau_{2}-t_{c}\right)\sum_{k}e^{-\nu_{k}\tau_{2}}\text{Tr}\left\{ \varrho_{k}^{(2)}\left(t-\tau_{2}-t_{c}\right)\right\} ,
\end{gather*}
with the Laplace image

\[
\tilde{p}_{{\rm f}}^{(2)}\left(s\right)=e^{-s\left(\tau_{2}+t_{c}\right)}\sum_{k}e^{-\nu_{k}\tau_{2}}\text{Tr}\left\{ \varrho_{k}^{(2)}\left(s\right)\right\} .
\]

Summing up the trajectories for the eventual successful generation
of the third segment we again obtain the infinite series, with the
Laplace image which converges to

\[
\tilde{\varrho}_{\text{prep},k}^{(2)}\left(s\right)\equiv\frac{\tilde{\varrho}_{{\rm nf},k}^{(2)}\left(s\right)}{1-\tilde{p}_{{\rm f}}^{(2)}\left(s\right)}.
\]

The successful, last,\textcolor{red}{{} }merging attempt of the third
segment $k$ with the rest of the state is represented as
\[
\Theta\left(t-t_{m}\right)e^{\mathcal{L}t_{m}}\sum_{k}\mathcal{\tilde{M}}{}_{k}\varrho_{\text{prep},k}^{(2)}\left(t\right),
\]
with $\mathcal{L}$ the superoperator describing degradation of the
final state during the time required by the merging. All possible
unsuccessful merging attempts are represented by 
\[
\Theta\left(t-t_{m}\right)\sum_{k}{\rm Tr}\left\{ \left(\mathcal{\mathcal{I}}-\mathcal{\tilde{M}}{}_{k}\right)\varrho_{\text{prep},k}^{(2)}(t)\right\} .
\]
Finally, in the Laplace domain, the density matrix distribution describing
the result of merging of three segments of the 2D network reads 
\begin{equation}
\tilde{\varrho}(s)=\frac{e^{\mathcal{L}t_{m}}\sum_{k}\mathcal{\tilde{M}}{}_{k}\tilde{\varrho}_{\text{prep},k}^{(2)}\left(s\right)}{e^{st_{m}}-\sum_{k}{\rm Tr}\left\{ \left(\mathcal{\mathcal{I}}-\mathcal{\tilde{M}}{}_{k}\right)\tilde{\varrho}_{\text{prep},k}^{(2)}\left(s\right)\right\} }.\label{eq:-10}
\end{equation}
The resulting Eq.~\eqref{eq:-10} allows us to find the average state
and the generation time of the larger segment as described in the
main text. Repeating this procedure recursively, we can simulate the
2D network of arbitrary nesting level depth. The results of the simulation
are presented the Figs.~\eqref{fig:-2}, and \eqref{fig:-2-1} of
the main text.

\section{Secret key rate\label{Appendix: Secret-key-rate}}

In the section, we review the formula for the rate of the $N$-partite
secret key \citet{Epping2017} generated out of $N$-partite entangled
state. The ideal state for the generation of the secret key among
$N$ parties, $A$ and $B_{i}$, with $i=1,\,..,\,N-1$, is the distributed
$N$-qubit GHZ state $\left\vert \text{GHZ}\right\rangle =(\left\vert 0\right\rangle _{\text{logic}}^{\otimes N}+\left\vert 1\right\rangle _{\text{logic}}^{\otimes N})/\sqrt{2}$,
encoded in the logical Hilbert space. To obtain the key, parties make
``first type'' or the ``second type'' measurement over the qubits.
The first type measurement is performed in the $Z$-basis $\left\{ \left\vert 0\right\rangle ,\left\vert 1\right\rangle \right\} _{\text{logic}}$,
and the second type measurement in $X$-basis $\{1/\sqrt{2}(\left\vert 0\right\rangle \pm\left\vert 1\right\rangle )_{\text{logic}}\}$
or $Y$-basis $\{1/\sqrt{2}(\left\vert 0\right\rangle \mp i\left\vert 1\right\rangle )_{\text{logic}}\}$
chosen randomly. The measurements of the second type allow calculation
of errors, potentially introduced by the eavesdropper. Based on the
measured errors, parties can reduce the length of the key by ``erasing''
the information that the eavesdropper could have obtained to make
the key completely secure. According to \citet{Lo2005} measurements
of the second type could be done much less frequently than of the
first, increasing the key distribution rate. At the end of the ``erasing''
step, all parties either have the unconditionally secure key or nothing,
depending on the distributed state quality.

Let us consider a quantum network which can generate $N$-partite
state $\rho$ for the average time $T$. Information in the state
is encoded in a logical subspace $\{\left\vert 0\right\rangle _{\text{logic}},\left\vert 1\right\rangle _{\text{logic}}\}$.
According to Ref.~\citet{Moroder2009}, without loss of security
one can discard all states that are out of the logical subspace. We
denote the projector to the logical subspace as $\Pi$. The probability
to find the network in the projected state, $\rho_{\Pi}\propto\Pi\rho\Pi$,
is $P_{\Pi}=\text{Tr}\left\{ \Pi\rho\Pi\right\} $. Then the secret
key rate could be calculated as
\[
R=r_{\infty}\left(\rho_{\Pi}\right)\frac{1}{T}P_{\Pi},
\]
where $r_{\infty}\left(\rho_{\Pi}\right)$ is the secret fraction,
i.e, the ratio of the number of secret bits and the number of shared
states $\rho_{\Pi}$ in the limit of the infinitely long key.

To derive $r_{\infty}\left(\rho_{\Pi}\right)$, in Ref.~\citet{Epping2017},
a depolarization procedure is introduced, which could be applied locally
to the $N$-qubit state, depolarizing it to a state diagonal in the
GHZ basis
\[
\left\vert \psi_{j}^{\pm}\right\rangle =\frac{1}{\sqrt{2}}\left(\left\vert 0\right\rangle \left\vert j\right\rangle \pm\left\vert 1\right\rangle \left\vert \bar{j}\right\rangle \right)_{\text{logic}}.
\]
In the state above, the first qubit belongs to the party $A$, $j$
takes the values $0,\,...,\,2^{N-1}-1$ in binary notation, and $\bar{j}$
denotes the binary negation of $j$; i.e., for example if $j=01101$
then $\bar{j}=10010$. Depolarized version of state $\rho_{\Pi}$
reads

\begin{multline}
\rho_{\Pi}^{\text{dep}}=\lambda_{0}^{+}\left\vert \psi_{0}^{+}\right\rangle \left\langle \psi_{0}^{+}\right\vert +\lambda_{0}^{-}\left\vert \psi_{0}^{-}\right\rangle \left\langle \psi_{0}^{-}\right\vert \\
+\sum_{j=1}^{2^{N-1}-1}\lambda_{j}\left(\left\vert \psi_{j}^{+}\right\rangle \left\langle \psi_{j}^{+}\right\vert +\left\vert \psi_{j}^{-}\right\rangle \left\langle \psi_{j}^{-}\right\vert \right),\label{eq:-17}
\end{multline}
where all coefficients could be found as $\left\langle \psi_{j}^{\pm}\right\vert \rho_{\Pi}\left\vert \psi_{j}^{\pm}\right\rangle .$ 

From decomposition~\eqref{eq:-17} one can find the probability that
the $Z$-measurement outcome of the party $B_{i}$ disagrees with
the one of the party $A$
\[
Q_{AB_{i}}=2\underset{j^{\left(i\right)}=1}{\sum_{j}}\lambda_{j},
\]
where $j^{\left(i\right)}$ denotes the $i^{\text{th}}$ bit of $j$.
The probabilities that at least one party $B_{i}$ obtains a different
outcome in the $Z$-bases than the party $A$ could be found as
\begin{align*}
Q_{Z} & =1-\lambda_{0}^{+}-\lambda_{0}^{-}.
\end{align*}

\noindent The probability that the $X$-measurement gives a result
incompatible with the noiseless state is
\[
Q_{X}=\frac{1-\lambda_{0}^{+}+\lambda_{0}^{-}}{2}.
\]

Finally, the secret fraction found un \citet{Epping2017} as
\begin{multline*}
r_{\infty}\left(\rho_{\Pi}\right)=\left(1-\frac{Q_{Z}}{2}-Q_{X}\right)\text{log}_{2}\left(1-\frac{Q_{Z}}{2}-Q_{X}\right)\\
+\left(Q_{X}-\frac{Q_{Z}}{2}\right)\text{log}_{2}\left(Q_{X}-\frac{Q_{Z}}{2}\right)\\
+\left(1-Q_{Z}\right)\left[1-\text{log}_{2}\left(1-Q_{Z}\right)\right]-h\left(\underset{1\leq i\leq N-1}{\text{max}}Q_{AB_{i}}\right),
\end{multline*}
with $h(p)=-p\,\text{log}_{2}p-\left(1-p\right)\text{log}_{2}\left(1-p\right)$
the binary Shannon entropy.

For example, Fig.~$\ref{fig:-2-1}$(a) of the main text presents
secret key distribution between two parties A and B with two parallel
DLCZ links, each of which has the target state $\left(\left\vert 0_{A}1_{B}\right\rangle +\left\vert 1_{A}0_{B}\right\rangle \right)/\sqrt{2}$
in the excitation encoding. The logical $Z$-basis for the measurement
of two local pares of the qubits at each party side is $\left\{ \left\vert 0\right\rangle ,\left\vert 1\right\rangle \right\} _{\text{logic}}^{A}=\{\left\vert 10\right\rangle _{A},\left\vert 01\right\rangle _{A}\}$
for the party A and $\left\{ \left\vert 0\right\rangle ,\left\vert 1\right\rangle \right\} _{\text{logic}}^{B}=\{\left\vert 01\right\rangle _{B},\left\vert 01\right\rangle _{B}\}$
for the party B, were the first and the second qubits of the states
belong to the first and the second links correspondingly. All multiple-click
or no-click events are discarded.

The tripartite key generation among parties A, B and C is shown in
Fig.~$\ref{fig:-2-1}$(b). The target state distributed with the
2D network is $\left(\left\vert 0_{A}0_{B}1_{C}\right\rangle +\left\vert 1_{A}1_{B}0_{C}\right\rangle \right)/\sqrt{2}$
in the excitation encoding. The logical basis for the parties A and
B is $\left\{ \left\vert 0\right\rangle ,\left\vert 1\right\rangle \right\} _{\text{logic}}^{A(B)}=\{\left\vert 0\right\rangle _{A(B)},\left\vert 1\right\rangle _{A(B)}\}$
and for the party C is $\left\{ \left\vert 0\right\rangle ,\left\vert 1\right\rangle \right\} _{\text{logic}}^{C}=\{\left\vert 1\right\rangle _{C},\left\vert 0\right\rangle _{C}\}$. 

In the paper, we consider the measurement of the memories states with
the imperfect photodetectors. Therefore, we apply the loss superoperator~$\eqref{eq:-28}$
$\mathcal{S}_{\text{loss}}\left(-\text{ln}[1-v]\right)$, with $v$
the detectors inefficiency, to each memory of the generated state
before the secret key is calculated.

\bibliography{bibtex}

\begin{thebibliography}{51}
\bibitem [{Par()}]{Parameters}%
  \BibitemOpen
  \href@noop {} {\bibinfo  {journal} {The following parameters were used:
  detection inefficiency $f=0.02$, read-out inefficiencies $v=0.02$, fiber
  attenuation length $L_\mathrm{att}=22$km, gate efficiency (for the 2D
  repeater only) $\eta=0.6$, signal length $l= 10^{-4}$s, and dark count
  probability in the signal measurement operation
  $d=10\mathrm{Hz}\cdot10^{-4}\mathrm{s}=0.001$}\ }\BibitemShut {NoStop}%
\end{thebibliography}

\end{document}